\documentclass[twoside,11pt]{article}
\usepackage[left=2cm,right=2cm,top=3cm,bottom=3cm]{geometry}
\usepackage{amsfonts,amsmath,mathtools,amssymb,dsfont}
\usepackage{tikz,pgfplots,graphicx}
\usepackage{accents}
\usepackage{authblk}
\usepackage[authoryear]{natbib}
\RequirePackage[colorlinks,citecolor=blue,urlcolor=blue,backref=page,backref=page]{hyperref}
\usetikzlibrary{shapes,decorations,arrows,calc,arrows.meta,fit,positioning}

\newcommand{\m}{\mathbf{m}}
\newcommand{\x}{\mathbf{x}}

\newcommand{\E}{\mathds{E}}
\newcommand{\N}{\mathrm{N}}
\newcommand{\w}{\mathrm{w}}
\newcommand{\iidsim}[0]{\stackrel{\mathrm{iid}}{\sim}}
\newcommand{\V}{ \mathbb{V}}

\newcommand{\Y}{\mathbf{Y}}
\newcommand{\ind}{\perp\!\!\!\!\perp} 
\DeclareMathSymbol{\shortminus}{\mathbin}{AMSa}{"39}
\newcommand{\ubar}[1]{\underaccent{\bar}{#1}}

\pgfplotsset{
	compat=1.11,
}
\pgfkeys{
	/pgf/declare function={f(\x,\a,\b) = \a*\x+\b;}
}
\pgfkeys{
	/pgf/declare function={g(\x,\a,\b) = \a*\x+\b;}
}
\newsavebox{\firsttree}
\newsavebox{\secondtree}
\newsavebox{\thirdtree}
\newsavebox{\fourthtree}
\newsavebox{\fifthtree}
\newsavebox{\firsttreeb}
\newsavebox{\secondtreeb}
\newsavebox{\thirdtreeb}
\newsavebox{\fourthtreeb}
\newsavebox{\firsttreec}
\newsavebox{\secondtreec}
\newsavebox{\thirdtreec}
\newsavebox{\fourthtreec}
\newsavebox{\fifthtreec}
\newsavebox{\partitionA}
\newsavebox{\partitionB}
\newsavebox{\partitionC}
\newsavebox{\treeA}
\newsavebox{\treeB}
\newsavebox{\compositeA}
\newsavebox{\compositeB}
\usepackage{subcaption}
\usepackage[edges]{forest}
\forestset{default preamble={for tree={draw}}}
\begin{document}

\title{A Bayesian Additive Regression Tree Model for Learning Conditional Average Treatment Effects in Regression Discontinuity Designs}

\author[1]{Rafael Alcantara\thanks{Rafael Alcantara gratefully acknowledges financial support from FAPESP grant 2021/13137-0 during his PhD in Business Economics at Insper under the supervision of Hedibert F. Lopes. Hedibert F. Lopes also acknowledges partial financial support from FAPESP grants 2023/02538-0 and 2024/01027-4.}}
\author[2]{P. Richard Hahn}
\author[3]{Hedibert F. Lopes}

\affil[1]{The University of Texas at Austin}
\affil[2]{Arizona State University}
\affil[3]{Insper}

\maketitle
		
		\begin{abstract}
			This paper develops an effective Bayesian approach to conditional average treatment effect (CATE) estimation in regression discontinuity designs (RDD), an increasingly prevalent form of quasi-experiment that facilitates causal inference. Earlier Bayesian approaches do not easily accommodate CATE estimation while recent frequentist approaches to this problem assume a known basis expansion, a steep model specification requirement that our approach avoids. 
			
			The new model is a variant of a Bayesian additive regression tree (BART) model with linear leaf-level regressions on the running variable and a treatment dummy (and their interaction). The model adaptively partitions covariate space into regions where the slope on the running variable appreciably differs, providing interpretable Bayesian inference on conditional average treatment effects near the cutoff. 
		\end{abstract}
		
		\paragraph{Keywords:} Bayesian additive regression trees, BART, CATE, conditional average treatment effects, RDD, regression discontinuity design
			
	\section{Introduction}
	\label{sec:introduction}
	Regression discontinuity designs (RDD), originally proposed by \citet{thistlethwaite1960regression}, are widely used in economics and other social sciences to estimate treatment effects from observational data. Such designs arise when treatment assignment is based on whether a particular covariate --- referred to as the running variable --- lies above or below a known value, referred to as the cutoff value. Because treatment is deterministically assigned as a known function of the running variable,  RDDs are trivially deconfounded: treatment assignment is independent of the outcome variable, given the running variable (because treatment is conditionally constant). 
	
	However, estimation of treatment effects in RDDs is more complicated than simply controlling for the running variable, because doing so introduces a complete lack of overlap, which is the other key condition needed to justify regression adjustment for causal inference. Nonetheless, treatment effects {\em at the cutoff} may still be identified.	Specifically, it is well-known that treatment effects at the cutoff can be estimated from RDDs as the magnitude of a discontinuity in the conditional mean response function at that point \citep{hahn2001identification}. 
	
	This paper investigates the use of Bayesian additive regression tree models \citep{chipman2010bart, hahn2020bayesian} for the purpose of estimating conditional average treatment effects (CATE) at the cutoff, conditional on observed covariates other than the running variable. More specifically, the new model is a variant of BART with linear leaf-level regressions on the running variable and a treatment dummy (and their interaction). The model adaptively partitions the covariate space into regions where the slope on the running variable appreciably differs, providing interpretable Bayesian inference on conditional average treatment effects near the cutoff. 
	
	The justification/motivation for developing this novel and non-trivial extension to BART is simply this: BART is widely-acknowledged to be superior at estimating conditional expectations compared to the methods that underpin previous RDD approaches to CATE estimation (linear models with pre-specified nonlinear basis functions or single individual regression tree models). The paper presents a detailed simulation experiment that bears this out.
	
	The simulation experiments described in section \ref{simulations} constitute another novel contribution of this paper. It is common practice in the methodological RDD literature to propose variants of a local (to the cutoff) polynomial model on the running variable and investigate its properties in data simulated with a polynomial mean function\footnote{A deeper look into the simulation studies in some of the most relevant RDD papers can be found at \url{https://github.com/rafaelcalcantara/BART-RDD/}}. Even though local polynomial models might provide good boundary point approximations asymptotically in many cases, the small sample quality of these approximations can vary substantively if the running variable affects the outcome non-linearly, or if it interacts with other covariates in complicated ways. The common practice might, therefore, provide an overly optimistic assessment of the local polynomial estimator in many cases. This problem is aggravated by the fact that many studies consider implausibly high signal levels in their simulations. For this reason, we move away from the DGPs commonly evaluated and provide a framework that:
	\begin{itemize}
		\item provides better control of DGP characteristics that are relevant for CATE estimation in RDD
		\item is not tailored to any particular modelling assumptions, which makes it a more general framework for comparing RDD CATE estimators
		\item considers lower signal-to-noise ratios, which are calibrated \textbf{at the cutoff}, as opposed to being defined globally
	\end{itemize}

	\subsection{Previous work}
	
	\subsubsection{Non-Bayesian work on CATE in RDDs}
	The inclusion of covariates in RDD models has been studied
	by a number of authors, but primarily from the
	perspective of obtaining precision gains for average treatment effect (ATE) estimation (at the cutoff), mostly in the context of linear models, and mostly from a frequentist perspective. 
	
	Two previous approaches to CATE estimation appear in earlier literature.
	
	\begin{itemize}
		\item	Both \cite{becker2013absorptive} and \cite{calonico2025treatment} extend the traditional
		local regression to include interaction terms between the
		treatment dummy and {\em known} smooth basis functions of additional
		covariates. By contrast, the method developed in this paper will not need to assume a pre-specified basis expansion, instead using a Bayesian tree-ensemble algorithm to facilitate effective data-driven CATE estimation. 
		
		\item 	\cite{reguly2021heterogeneous} proposes a
		modified CART (classification and regression tree) algorithm in which the tree is split using all
		features available {\em except} for the running variable; then,
		within each leaf the algorithm performs a separate
		regression for treated and untreated units, and the
		leaf-specific ATE parameter is obtained as the difference
		between the intercepts of the two regressions. The many ways that the approach developed in this paper differs from \cite{reguly2021heterogeneous} will be revisited after the new approach has been spelled out in detail.
		
	\end{itemize}

	\subsubsection{Bayesian RDD work}
	A prominent recent example of a Bayesian estimator for RDDs is
	\cite{chib2023nonparametric}, who estimate the response curves
	with global splines where observations are weighted by their
	distance to the cutoff. However, the assumptions underlying this model permit ATE estimation, but not CATE estimation. 
	
	Two other notable works are \cite{karabatsos2015bayesian}, who propose approximating the conditional expectations by an
	infinite mixture of normals and \cite{branson2019nonparametric}, who propose a Gaussian
	process prior for the expectations. While these approaches could perform CATE estimation in principle, these authors do not evaluate their models on CATE estimation (focusing instead on the ATE). Moreover, their steep computational demands make thorough CATE comparisons infeasible for larger sample sizes (which are necessary for effective CATE estimation in practice).
	
	Other related Bayesian work includes \cite{sugasawa2023hierarchical} and \cite{tao2025bayesian}, who propose Bayesian hierarchical models for heterogeneous effect estimation, where a local linear and a Gaussian process regression (respectively) are considered for each subgroup. These methods require \textit{a priori} knowledge of the subgroups that define heterogeneity and rely on restrictive parametric assumptions. In contrast, adapting BART to the RDD context allows us to learn effect heterogeneity in a data-driven manner for more complex response surfaces.

	\subsubsection{Extensions to BART}
	Our work is also a contribution to the burgeoning field of extensions to BART. BART models have been greatly extended in the years since 2010, to include heteroskedastic variants \citep{pratola2020heteroscedastic, murray2021log}, classification \citep{murray2021log}, conditional density estimation \citep{orlandi2021density}, variable selection \citep{linero2018bayesianDART}, monotonicity constraints \citep{chipman2022mbart, papakostas2023forecasts}, survival analysis \citep{sparapani2016nonparametric}, partial identification \citep{hahn2016bayesian} among others. 
	
	Specifically, this paper is a contribution to the literature on BART for causal inference  \citep{hill2011bayesian,hahn2020bayesian}, adapting these earlier works to the unique problems posed by regression discontinuity designs. As was the case in \cite{hahn2020bayesian}, direct application of BART to treatment effect estimation turns out to be suboptimal and certain causal-inference-specific model adjustments are proposed, which we turn to next.
	
	\subsection{BART for Causal Inference}
	As in regression adjustments for treatment effect estimation under strong ignorability \citep{hill2011bayesian,hahn2020bayesian}, conditional average treatment effect estimation in RDDs boils down to estimating (potentially complex) conditional expectation functions (see Sections \ref{sec:rdd} and \ref{sec:cate_id}). BART may be used to fit the conditional expectations needed for treatment effect estimation in RDDs in a number of different ways. The first way, what is sometimes called an ``S-Learner'',  is simply to include the treatment assignment indicator among the feature set along with the other covariates. This approach was proposed in \cite{hill2011bayesian} in the context of regression adjustment for treatment effect estimation under conditional strong ignorability. The second way, what is sometimes called a ``T-Learner'' is to fit two individual BART models to the treated and control data separately. See \cite{kunzel2019metalearners} for a discussion of the S- and T-Learner nomenclature. 
	
	\cite{hahn2020bayesian} provide an extensive discussion of potential drawbacks to the S-Learner and T-Learner approaches in the context of BART models, which we now summarize. We will refer to S-BART and T-BART in relation to these strategies. The main problem with S-BART is that there can be many trees, with potentially very different splits, that achieve a similar likelihood evaluation. While this is unobjectionable if the goal is merely to predict the response surface -- indeed, the over-parametrization of these models probably accounts for much of their empirical success --- different splitting patterns in treatment assignment versus other covariates tend to imply distinctly different treatment effect estimates. The upshot --- borne out by extensive simulation studies --- is that S-BART models for causal inference tend to have unpredictable biases as a result of the specific dependency structure among the predictor variables in a given data set. While the T-BART approach successfully addresses this drawback of S-BART, fitting completely separate models to the treated and control data introduces a different problem: regularization of the conditional treatment effect function itself is implicit, and generally too weak when treatment effects are expected to be small relative to variation in the outcome due to other observed features or unobserved factors. In short, T-BART tends to over-fit the data, yielding CATE estimators with high variance.
	
	In Section \ref{sec:barddt}, we introduce a modified BART model --- different from both S-BART and T-BART --- that is markedly better at CATE estimation than either one, and also better than local polynomial regression and the CART approach of \cite{reguly2021heterogeneous} as well. The new model uses a linear regression leaf model within a BART ensemble, fitted with software developed by our lab for this purpose.
\section{Background}
To keep the paper relatively self-contained, we briefly review the basics of regression discontinuity designs and BART, and cast the RDD problem from a functional causal model perspective that is convenient for BART modeling. 

\subsection{Regression Discontinuity Designs}
\label{sec:rdd}

We conceptualize the treatment effect estimation problem via a quartet of random variables $(Y, X, Z, U)$. The variable $Y$ is the outcome variable; the variable $X$ is the running variable; the variable $Z$ is the treatment assignment indicator variable; and the variable $U$ represents additional, possibly unobserved, causal factors. We suppose that realizations of $(Y_i, X_i, Z_i)$ for $i = 1, \dots, n$ are independent and identically distributed. What specifically makes this correspond to an RDD is that we stipulate that $Z = \mathbb{I}(X > c)$, for cutoff $c$\footnote{Treatment could be assigned to all units below the cutoff, rather than above, but this is irrelevant for identification, so we define treatment for units above the cutoff without loss of generality}. More precisely, this corresponds to a \textbf{sharp} RDD, which is the main focus of this paper\footnote{The other form of RDD commonly used in applied literature is the so-called \textbf{fuzzy} RDD, in which treatment is assigned with some probability $0 < P(Z=1 \mid X=c) < 1$ for units above the cutoff}. For the remainder of this paper we assume $c = 0$ without loss of generality.  

We may express $Y$ as some function of the random variables $(X,Z,U)$: $$Y = F(X,Z,U).$$ In principle, we may obtain draws of $Y$ by first drawing $(X,Z,U)$ according to their joint distribution and then applying the function $F$. Similarly, we may relate this formulation to the potential outcomes framework straightforwardly: 	
\begin{equation}
	\begin{split}
		Y^1 &= F(X,1,U),\\
		Y^0 &= F(X,0,U).
	\end{split}
\end{equation}
Here, draws of $(Y^1, Y^0)$ may be obtained (in principle) by drawing $(X,Z,U)$ from their joint distribution and using only the $(X,U)$ elements as arguments in the above two equations, ``discarding'' the drawn value of $Z$. Note that this construction implies the {\em consistency} condition: $Y = Y^1 Z + Y^0 ( 1 - Z)$. Likewise, this construction implies the {\em no interference} condition because each $Y_i$ is considered to be produced with arguments ($X_i, Z_i, U_i)$ and not those from other units $j$; in particular, in constructing $Y_i$, $F$ does not take $Z_j$ for $j \neq i$ as an argument.

Next, we define the following conditional expectations
\begin{equation}
	\begin{split}
		\mu_1(x) &= \E[ F(x, 1, U) \mid X = x] \\
		\mu_0(x) &= \E[ F(x, 0, U) \mid X = x],
	\end{split}
\end{equation}
with which we can define the treatment effect function
$$\tau(x) = \mu_1(x) - \mu_0(x).$$

In potential outcomes terminology, the deterministic treatment assignment mechanism in a sharp RDD implies that conditioning on $X$ satisfies ignorability,
$$(Y^1, Y^0) \ind Z \mid X,$$
but not {\em strong ignorability}, because overlap is violated, as that would require that
$$0 < \mathbb{P}(Z = 1 \mid X=x) < 1 \;\;\;\; \forall x.$$
Therefore, we can in fact only learn $\mu_1(x)$ for $x > 0$ and $\mu_0(x)$ for $x \leq 0$. That is:
\begin{equation} \label{eq:potential.outcomes}
	\begin{split}
		\mu_1(x) &= \E[Y \mid X=x, Z=1] \quad \text{for } x>0,\\
		\mu_0(x) &= \E[Y \mid X=x, Z=0] \quad \text{for } x \leq 0,
	\end{split}
\end{equation}	
the right-hand-sides of which can be estimated from sample data.
Because we can only learn $\mu_0(x)$ or $\mu_1(x)$ for given $x$, $\tau(x)$ is unidentified without further assumptions. However, it is possible to learn $\tau(0)$ for continuous $X$ so long as one is willing to assume that $\mu_1(x)$ and $\mu_0(x)$ are both suitably smooth functions of $x$: any inferred discontinuity at $x = 0$ must therefore be attributable to treatment effect\footnote{See \cite{hahn2001identification} for the seminal exposition of continuity-based identification in RDD from the potential outcomes perspective}. Formally, we assume that $\mu_1,\mu_0$ are continuous at $x=0$, \textit{i.e.}:
\begin{equation}
	\begin{split}
		\lim_{x \to 0^+} \mu_1(x) &= \lim_{x \to 0^-} \mu_1(x) = \mu_1(0)\\
		\lim_{x \to 0^+} \mu_0(x) &= \lim_{x \to 0^-} \mu_0(x) = \mu_0(0).
	\end{split}
\end{equation}
This implies that we can estimate $\tau(0)$ by learning $\E[Y \mid X=x,Z=1],\E[Y \mid X=x,Z=0]$ and predicting each of those functions at $x=0$. Importantly, note that our target estimand is simply $\tau(0)$, which is not defined by the limits of $\mu_1,\mu_0$; these limits only matter insofar as continuity of these functions leads to identification of $\tau(0)$.

\subsubsection{Conditional average treatment effects in RDD}\label{sec:cate_id}

In this paper, we are concerned with learning not only $\tau(0)$, the “RDD ATE” (e.g. the CATE at $x = 0$), but also RDD CATEs, $\tau(0, \w)$ for some covariate vector $\w$. Incorporating additional covariates in the above framework turns out to be straightforward, simply by defining $W = \varphi(U)$ to be an observable function of the (possibly unobservable) causal factors $U$. We may then define our potential outcome
means as
\begin{equation}
	\begin{split}
		\mu_1(x, \w) &= E[F(x, 1, U) | X = x, W = \w]\\
		\mu_0(x, \w) &= E[F(x, 0, U) | X = x, W = \w],
	\end{split}
\end{equation}
and our treatment effect function as
\begin{equation}
	\tau(x,\w) = \mu_1(x,\w) - \mu_0(x,\w).
\end{equation}
We consider our data to be independent and identically distributed realizations $(Y_i,X_i,Z_i,W_i)$ for $i =1, . . . , n$.

To identify $\tau(0,\w)$ we require that, for each covariate value $\w$ of interest, the distribution of $(X,W)$ places positive probability arbitrarily close to $(0,\w)$ on each side of the cutoff. For discrete covariates, “close to $\w$” means equal to $\w$. Also, we require that:

\paragraph{For discrete $W$}
For all $\w$:
\begin{equation}
	\begin{split}
		\lim_{x \to 0^+} \mu_1(x,\w) &= \lim_{x \to 0^-} \mu_1(x,\w) = \mu_1(0,\w)\\
		\lim_{x \to 0^+} \mu_0(x,\w) &= \lim_{x \to 0^-} \mu_0(x,\w) = \mu_0(0,\w).
	\end{split}
\end{equation}
\paragraph{For continuous $W$} $\mu_1$ and $\mu_0$ are jointly continuous in $(x,\w)$ at $x=0$.
\hfill \break

Under these assumptions, $\mu_1(x,\w)$ and $\mu_0(x,\w)$ are estimable for data near $x=0$:
\begin{equation} \label{eq:estimable}
	\begin{split}
		\mu_1(x,\w) &= \E[Y \mid X=x, W=\w, Z=1] \quad \text{for } x>0,\\
		\mu_0(x,\w) &= \E[Y \mid X=x, W = \w, Z=0] \quad \text{for } x \leq 0,
	\end{split}
\end{equation}
and $\tau(0,\w)$ is identified by the difference of two estimable conditional expectations:
\begin{equation}
	\tau(0,\w) = \lim_{x \to 0^+} \E[Y \mid X=x,W=\w,Z=1]-\E[Y \mid X=0,W=\w,Z=0].
\end{equation}
Thus, CATE estimation in RDDs boils down to estimation of conditional expectation functions $\E[Y \mid X=x, W=\w, Z=z]$, for which we turn to BART models.

	\subsection{Bayesian Additive Regression Trees}
	\label{sec:bart}
	The Bayesian Additive Regression Trees model
	\citep{chipman2010bart}, or BART, represents an unknown mean
	function as a sum of regression trees, where each regression
	tree is assigned the prior described in
	\cite{chipman1998bayesian}. In this section we describe the model in terms of generic predictor vector $X$, to match earlier work, but in subsequent sections we will specialize our notation to include $(X, W, Z)$ as in the RDD notation of the previous section.

	Letting \(f(x) = \E(Y \mid X = x)\)
	denote the unknown mean function of a covariate vector \(X\), a BART
	model with $J$ trees is traditionally written
	\begin{equation} \label{eq:bart.model}
		\begin{split}
			Y &= f(x) + \varepsilon,\\
			&= \sum_{j=1}^J g(x; T_j, \m_j) + \varepsilon,\\
			& = \sum_{j=1}^J g_j(x) + \varepsilon,
		\end{split}
	\end{equation}
	where \(\varepsilon \sim \N(0,\sigma^2)\) is a normally
	distributed additive error term. Here, \(g(x ; T_j,
	\m_j)\) denotes a piecewise constant function of \(x\) defined by a set
	of splitting rules \(T_j\) that partition the domain $\mathcal{X}$
	into $B_j$ disjoint regions, and a vector, $\m_j = (m_{j,1}, \dots, m_{j,B_j})$, which records
	the values taken by \(g(\cdot)\) on each of those
	regions. That is, let $b_j(x): \mathcal{X} \rightarrow \lbrace 1, \dots, |\m_j|  = B_j \rbrace$ be a function denoting which leaf node of the $j$th tree contains the point $x$; then 
	$$g_j(x) = g(x; T_j, \m_j) =	m_{j,b_j(x)}.
	$$
	Therefore, the parameters of a standard BART
	regression model are \((T_1, \m_1), \dots, (T_J, \m_J)\) and
	\(\sigma\).  \cite{chipman2010bart} consider priors such that:
	the tree components \((T_j, \m_j)\) are independent of each
	other and of \(\sigma^2\), and the leaf node parameters
	$m_{j,b}$ are all mutually
	independent. Furthermore, \cite{chipman2010bart}
	specify the same priors for all trees and leaf node
	parameters. The model thus consists of the specification of
	three priors: \(p(T)\), \(p(\sigma^2)\) and \(p(\m \mid T)\).
	
	The tree prior, \(p(T)\), is defined by three
	components. First, the probability that a node with depth \(d\) will
	split is 
	\begin{equation}
		\frac{\alpha}{(1+d)^\beta}, \quad \alpha \in (0,1), \beta \in [0,\infty), \label{eq:tree.prior.1}
	\end{equation}
	implying that trees of greater depth have lower prior probability. The prior over cutpoints of the regression trees is uniform on the observed range of each feature and each feature is given equal prior weight. For the prior on the leaf node parameters, \(p(\m \mid T)\), \cite{chipman2010bart} specify independent Gaussian distributions over the elements of the $\m_j$ vectors: $m_{j,b} \iidsim \N(m_0,\sigma_0^2)\). Finally, $\sigma^2$ is given an inverse Gamma prior. For further details and justification concerning BART prior specification, see \cite{chipman2010bart}.
	
	\cite{chipman2010bart} construct a Gibbs sampling algorithm to obtain posterior draws of the trees, their leaf parameters, and the residual variance $\sigma^2$. Let $T_{\shortminus j}$ denote the set of all trees {\em except} $T_j$, and similarly for $\m_{\shortminus j}$. At each iteration, the algorithm produces $J$ consecutive samples of $(T_j, \m_j, \sigma)$ using the following compositions:
	
	\begin{equation}
		T_j \mid T_{\shortminus j}, \m_{\shortminus j},\sigma,y, \label{eq:tree.gibbs}
	\end{equation}
	then
	\begin{equation}
		\m_j \mid T_j, T_{\shortminus j}, \m_{\shortminus j},\sigma,y, \label{eq:mu.gibbs}
	\end{equation}
	and finally
	\begin{equation}
		\sigma \mid T_1,\m_1, \dots T_J, \m_J,y.
	\end{equation}
	Sampling from \eqref{eq:tree.gibbs} is simplified by noting that each tree depends on $(T_{\shortminus j}, \m_{\shortminus j},y)$ only through a ``partial'' residual:
	\begin{equation}
		\mathrm{r}_j = y - \sum_{j' \neq j} g(x;T_{j'},\m_{j'}).\label{eq:partial.residual}
	\end{equation}
	Therefore, the log marginal likelihood of the data in leaf $b_j$ (integrating out the unknown leaf mean parameter $m_{j, b_j}$) is:
	\begin{equation}
		l_{b_j} = -\frac{n_{b_j}}{2} \log(2\pi) - n_{b_j} \log(\sigma) + \frac{1}{2} \log \left( \frac{\sigma^2}{n_{b_j} \sigma^2_{\mu} + \sigma^2} \right) - \frac{\sum_{i: x_i \in b_j} r_i^2}{2 \sigma^2} + \frac{\sigma_{\mu}^2 (\sum_{i: x_i \in b_j} r_i)^2}{2\sigma^2(n_{b_j} \sigma_{\mu}^2 + \sigma^2)}, \label{eq:likelihood.bart}
	\end{equation}
	which is used to compute a Metropolis-Hastings ratio for accepting or rejecting a proposed tree. Details may be found in \cite{chipman2010bart} . Conditional on the tree, sampling the elements of $\m_j$ (step \ref{eq:mu.gibbs}) is a standard conjugate update that can be found in any textbook (see, for example, section 2.3 in \citet{gamerman2006markov}), where the observed ``data'' is the $r_j$ vector from just above. 
	
	\section{Bayesian Additive RDD Trees}\label{sec:barddt}
	For RDD, we propose that a linear model in the leaf is a viable strategy for overcoming the problems with T-BART and S-BART described above. We build on the work of \cite{chipman2002bayesian},  \cite{gramacy2008bayesian}, and \cite{starling2020bart}, by proposing a BART model where the trees are allowed to split on $(x,\w)$ but where each leaf node parameter is a vector of regression coefficients tailored to the RDD context (rather than a scalar constant as in default BART). In one sense, such a model can be seen as implying distinct RDD ATE regressions for each subgroup determined by a given tree; however, this intuition is only heuristic, as the entire model is fit jointly as an ensemble of such trees. Instead, we motivate this model as a way to estimate the necessary conditional expectations via a parametrization where the conditional treatment effect function can be explicitly regularized, as follows.
		
	Let $\psi$ denote the following basis vector:
	\begin{equation}
		\psi(x,z) = \begin{bmatrix}
			1 & z x & (1-z) x & z
		\end{bmatrix}.
	\end{equation}
	To generalize the original BART model, we define $g_j(x, \w, z)$ as a piecewise linear function as follows.  Let $b_j(x, \w)$ denote the node in the $j$th tree which contains the point $(x, \w)$; then the prediction function for tree $j$ is defined to be:
	\begin{equation}
		g_j(x, \w, z) = \psi(x, z) \Gamma_{b_j(x, \w)}
	\end{equation}	
	for a leaf-specific regression vector $\Gamma_{b_j} = (\eta_{b_j}, \lambda_{b_j}, \theta_{b_j}, \Delta_{b_j})^t$. Therefore, letting $n_{b_j}$ denote the number of data points allocated to node $b$ in the $j$th tree and $\Psi_{b_j}$ denote the $n_{b_j} \times 4$ matrix, with rows equal to $\psi(x,z)$ for all $(x_i,z_i) \in b_j$, the model for observations assigned to leaf $b_j$, can be expressed in matrix notation as:
	\begin{equation}
		\begin{split}
			\Y_{b_j} \mid \Gamma_{b_j}, \sigma^2 &\sim \N(\Psi_{b_j} \Gamma_{b_j},\sigma^2)\\
			\Gamma_{b_j} &\sim \N (0, \Sigma_0),
		\end{split} \label{eq:leaf.regression}
	\end{equation}
	where we set $\Sigma_0 = \frac{0.1}{J} \mbox{I}$ as a default (for $x$ vectors standardized to have unit variance in-sample)\footnote{We find that the particular choice for the leaf coefficient scale parameter does not typically matter in terms of performance of our model, except for when the scale is set unreasonably low. In particular, we ran our model for 2 of the simulation setups we analyze in section \ref{simulations} considering alternative scale values of $$\begin{bmatrix}
			\frac{0.01}{J}I & \frac{0.05}{J}I & \frac{0.5}{J}I & \frac{1}{J}I
		\end{bmatrix}.$$ The results are only qualitatively different for the $\frac{0.01}{J}$ case. The results of this analysis are available in an appendix}.
	
	This choice of basis entails that the RDD CATE at $\w$,  $\tau(0, \w)$, is a sum of the $\Delta_{b_j(0, \w)}$ elements across all trees $j = 1, \dots, J$:
	
	\begin{equation}
		\begin{split}
			\tau(0, \w) &= \E[Y^1 \mid X=0, W = \w] - \E[Y^0 \mid X = 0, W = \w]\\
			& =  \E[Y \mid X=0, W = \w, Z = 1] - \E[Y \mid X = 0, W = \w, Z = 0]\\
			&=  \sum_{j = 1}^J g_j(0, \w, 1) -  \sum_{j = 1}^J g_j(0, \w, 0)\\
			&= \sum_{j = 1}^J \psi(0, 1) \Gamma_{b_j(0, \w)}  - \sum_{j = 1}^J \psi(0, 0) \Gamma_{b_j(0, \w)} \\
			& = \sum_{j = 1}^J  \Bigl( \psi(0, 1) - \psi(0, 0) \Bigr)  \Gamma_{b_j(0, \w)} \\
			& = \sum_{j = 1}^J  \Bigl( (1,0,0,1) - (1,0,0,0)  \Bigr)  \Gamma_{b_j(0, \w)} \\
			&= \sum_{j=1}^J \Delta_{b_j(0, \w)}.
		\end{split}
	\end{equation}
	As a result, the priors on the $\Delta$ coefficients directly regularize the treatment effect. We set the tree and error variance priors as in the original BART model. 
	
	Posterior sampling from this model proceeds nearly identically to the traditional BART Gibbs sampler, but with a modified log marginal likelihood, which for a node $b_j$ is:
	\begin{equation}
		\begin{split}
			l_{b_j} = &-\frac{n_{b_j}}{2} \log(2\pi) - n_{b_j} \log(\sigma) - \frac{1}{2} \log \left( \det \left( \mbox{I} + \frac{\Sigma_0 \Psi_{b_j}^t\Psi_{b_j}}{\sigma^2} \right) \right)\\
			&- \frac{r_{b_j}^t r_{b_j}}{2 \sigma^2} + \frac{1}{2} \frac{r_{b_j}^t \Psi_{b_j}}{\sigma^2} \left( \Sigma_0^{-1} + \frac{\Psi_{b_j}^t \Psi_{b_j}}{\sigma^2} \right)^{-1} \frac{\Psi_{b_j}^t r_{b_j}}{\sigma^2}, \label{eq:likelihood.barddt}
		\end{split}
	\end{equation}
	where $r_{b_j}$ is a $n_{b_j}$ vector containing the partial residuals, as defined in \eqref{eq:partial.residual}, for the points in $b_j$. Note that this expression generalizes \eqref{eq:likelihood.bart}; the two expressions become equivalent if the basis vector $\Psi_{b_j}$ in the above expression is a single column of ones.
	
	Likewise, the parameter sampling follows a standard conditionally (on $\sigma^2$) conjugate linear regression update, independently for each leaf of the current tree which we omit here as it can be found in standard references (for example, section 2.3.3 in \citet{gamerman2006markov}). 
	
\begin{figure}
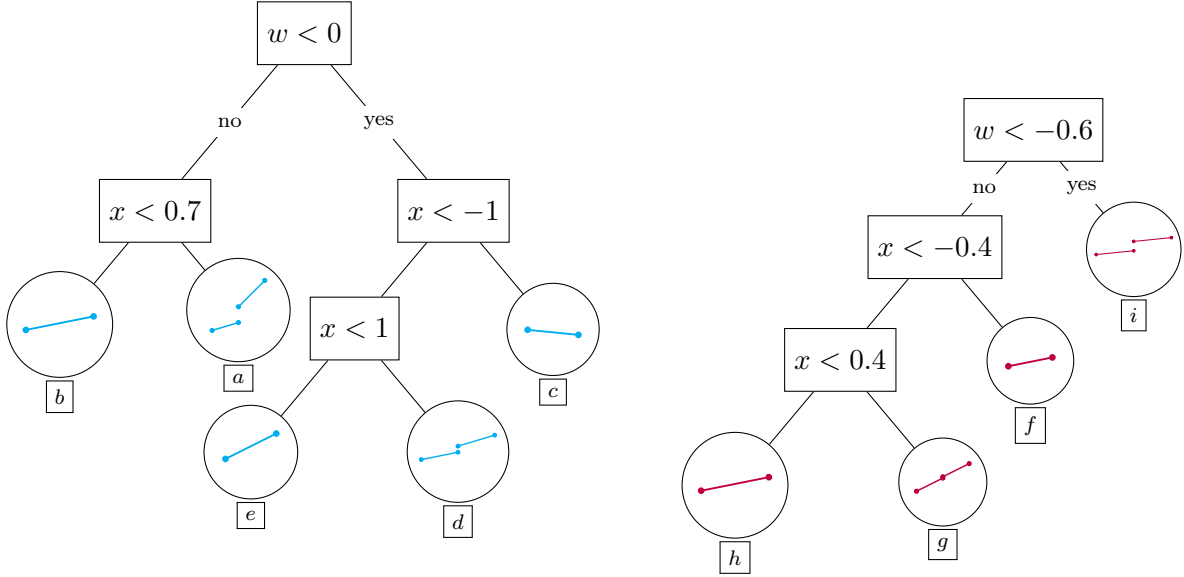

		\begin{subfigure}{0.5\textwidth}
			\usebox{\treeA}
			\label{fig:tree.1}
		\end{subfigure}
		\hfil\begin{subfigure}{0.45\textwidth}
			\usebox{\treeB}
			\label{fig:tree.2}
		\end{subfigure}
		\caption{\small Two regression trees with splits in $x$ and a single scalar $w$. Node images depict the $g(x,\w,z)$ function (in $x$) defined by that node's $\Gamma$ coefficients. The vertical gap between the two line segments in a node that contain $x=0$ is that node's contribution to the CATE at $X = 0$. Note that only such nodes contribute for CATE prediction at $x=0$} \label{cartoon1}
		\end{figure}

\begin{figure}
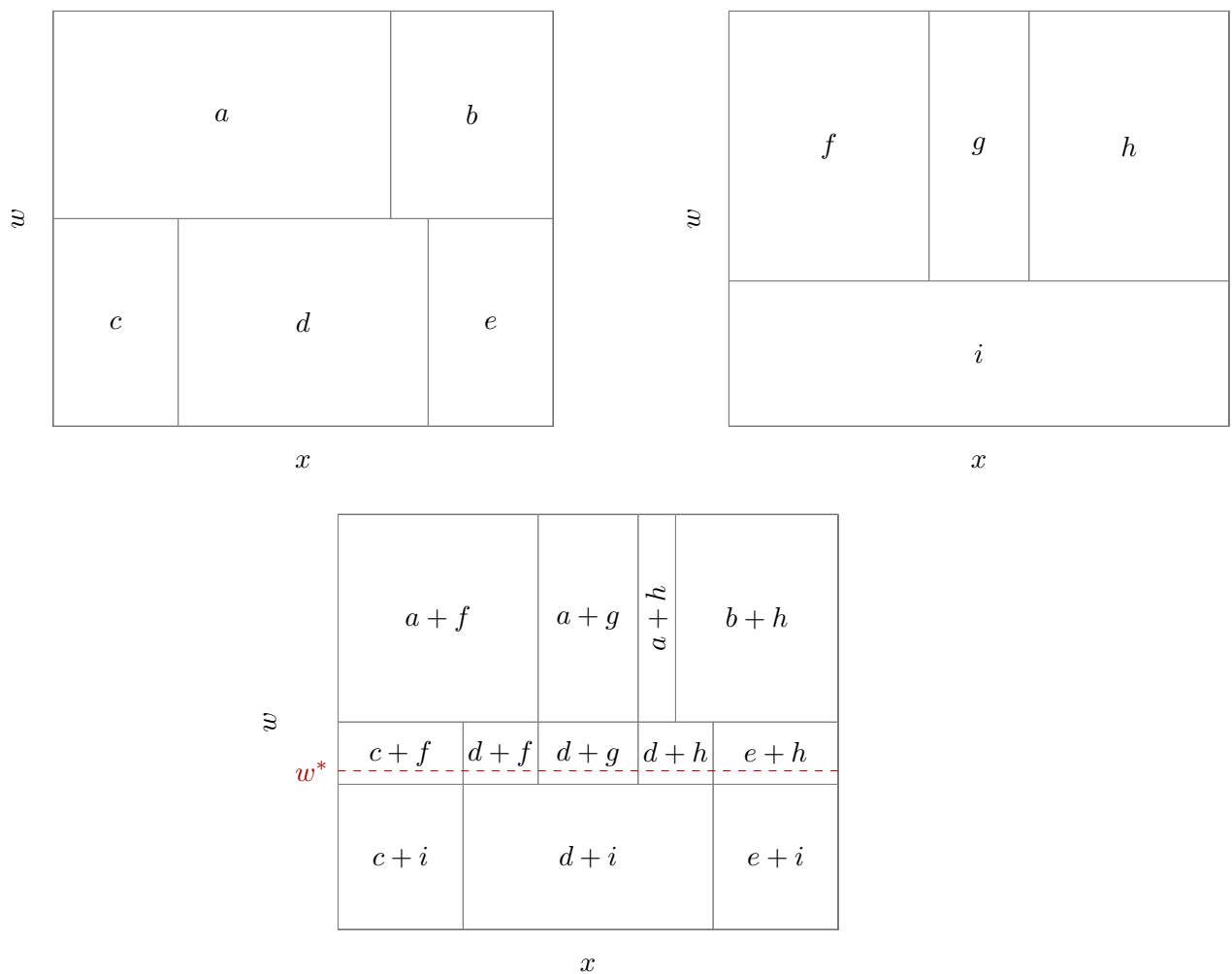

		\begin{subfigure}{0.9\linewidth}
			\centering
			\usebox{\partitionA}
			\label{fig:part.1}
		\end{subfigure}
		\hfil\hfil\hfil
		\begin{subfigure}{0.9\linewidth}
			\centering
			\usebox{\partitionB}
			\label{fig:part.2}
		\end{subfigure}
		\hfil\hfil\hfil
		\begin{subfigure}{0.9\linewidth}
			\centering
			\usebox{\partitionC}
			\label{fig:part.3}
		\end{subfigure}
		\caption{\small The two top figures show the same two regression trees as in the preceding figure, now represented as a partition of the $x\mbox{-}w$ plane. Labels in each partition correspond to the leaf nodes depicted in the previous picture. The bottom figure shows the partition of the $x\mbox{-}w$ plane implied by the sum of the two trees; the red dashed line marks point $W=w^*$ and the combination of nodes that include this point} \label{cartoon2}

		\end{figure}

\begin{figure}
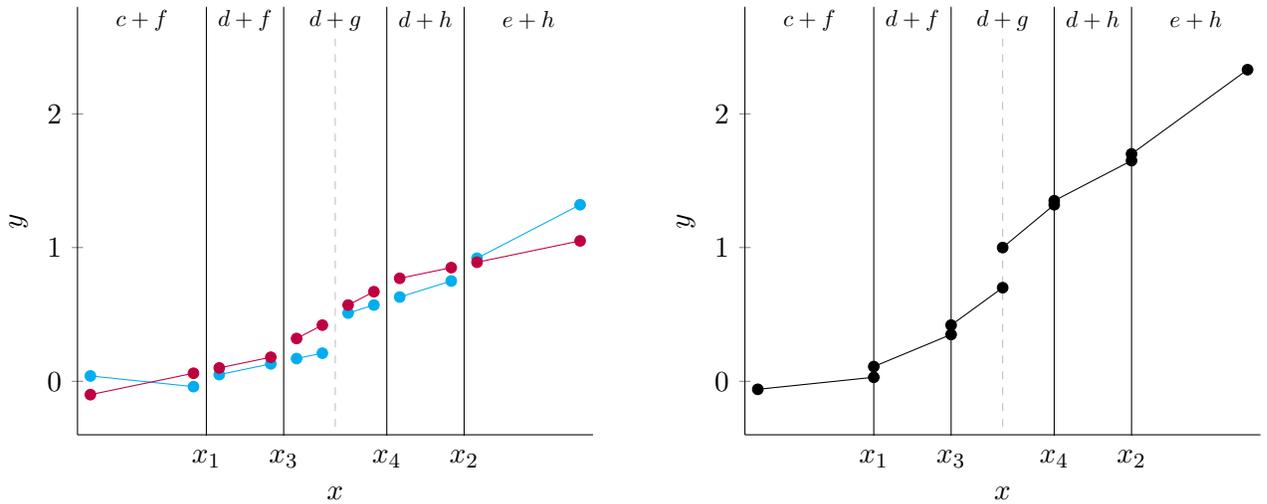

		\begin{subfigure}{0.5\linewidth}
			\centering
			\usebox{\compositeA}
			\label{fig:composite.1}
		\end{subfigure}
		\hfil\hfil\hfil
		\begin{subfigure}{0.5\linewidth}
			\centering
			\usebox{\compositeB}
			\label{fig:composite.2}
		\end{subfigure}
		\caption{\small Top: The function fit at $W = w^*$ for the two trees shown in the previous two figures, shown superimposed. Bottom: The aggregated fit achieved by summing the contributions of two regression tree fits shown at left. The magnitude of the discontinuity at $x = 0$ (located at the dashed gray vertical line) represents the treatment effect at that point. Different values of $w$ will produce distinct fits; for the two trees shown, there can be three distinct fits based on the value of $w$.} \label{cartoon3}
	\end{figure}

	Figures \ref{cartoon1} through \ref{cartoon3} provide a graphical depiction of how the BARDDT model fits a response surface and thereby estimates CATEs for distinct values of $\w$. For simplicity only two trees are used in the illustration, while in practice dozens or hundreds of trees may be used (in our simulations and empirical example, we use 50 trees). 
	
	An interesting property of BARDDT can be seen in this small illustration --- by letting the regression trees split on the running variable, there is no need to separately define a ``bandwidth'' as is used in the polynomial approach to RDD. Instead, the regression trees automatically determine (in the course of posterior sampling) when to ``prune'' away regions away from the cutoff value. There are two notable features of this approach. One, different trees in the ensemble are effectively using different local bandwidths and these fits are then blended together. For example, in the bottom panel of figure \ref{cartoon2}, we obtain one bandwidth for the region $d+i$, and a different one for regions $a+g$ and $d+g$. Two, for cells in the tree partition that do not span the cutoff, the regression within that partition contains no causal contrasts --- all observations either have $Z = 1$ or $Z = 0$. For those cells, the treatment effect coefficient is ill-posed and in those cases the posterior sampling is effectively a draw from the prior; however, such draws correspond to points where the treatment effect is unidentified and none of these draws contribute to the estimation of $\tau(0, \w)$ --- for example, only nodes $a+g$, $d+g$, and $d+i$ in figure \ref{cartoon2} provide any contribution. This implies that draws of $\Delta$ corresponding to nodes not predicting at $X=0$ will always be draws from the prior, which has some intuitive appeal.
	
	BARDDT differs from \citet{reguly2021heterogeneous} --- which is, to the best of our knowledge, the only other tree-based CATE estimator for RDD --- in three important ways:
	\begin{itemize}
		\item BARDDT is a sum of many regression trees, rather than a single tree
		\item the BARDDT estimator is based on Bayesian posterior mean\footnote{As a partition model, BART-based estimates of conditional expectations have points of discontinuity. Although RDD demands that $\mu_1$ and $\mu_0$ are smooth functions, this identification condition is on the DGP, not on the estimator. BART is a consistent estimator of that underlying smooth function \citep{he2023stochastic,saha2023theory} even if its estimates are not} rather than a single optimization-based model fit		
		\item BARDDT trees are permitted to split in the running variable (as mentioned in the previous paragraph)
	\end{itemize}
	
	\begin{lrbox}{\firsttree}{%
		\begin{tikzpicture}
			\draw[scale=0.5, variable=\x,color=cyan,thick] plot [mark=otimes*,samples at = {-1.5,0}]
			({\x}, {f(\x,-0.1,-0.15)});
		\end{tikzpicture}%
	}
\end{lrbox}
\begin{lrbox}{\secondtree}{%
		\begin{tikzpicture}
			\draw[scale=0.5, variable=\x,color=cyan,thick] plot [mark=otimes*, mark color=cyan,samples at = {-1.5,0}]
			({\x}, {f(\x,0.2,0.23)});
			\draw[-,scale=0.5, domain=0:1.5, variable=\x,color=cyan,thick] plot [mark=otimes*, mark color=cyan,samples at = {0,1.5}]
			({\x}, {g(\x,0.3,0.48)});
		\end{tikzpicture}%
	}
\end{lrbox}
\begin{lrbox}{\thirdtree}{%
		\begin{tikzpicture}
			\draw[-,scale=0.5, domain=0:1.5, variable=\x,color=cyan,thick] plot [mark=otimes*, mark color=cyan,samples at = {0,1.5}]
			({\x}, {g(\x,0.5,0.37)});
		\end{tikzpicture}%
	}
\end{lrbox}
\begin{lrbox}{\fourthtree}{%
		\begin{tikzpicture}
			\draw[-,scale=0.5, variable=\x,color=cyan,thick] plot [mark=otimes*, mark color=cyan,samples at = {-1,0}]
			({\x}, {f(\x,0.3,0.4)});
			\draw[-,scale=0.5, variable=\x,color=cyan,thick] plot [mark=otimes*, mark color=cyan,samples at = {0,1}]
			({\x}, {g(\x,1,1)});
		\end{tikzpicture}%
	}
\end{lrbox}
\begin{lrbox}{\fifthtree}{%
		\begin{tikzpicture}
			\draw[-,scale=0.5, domain=0:2, variable=\x,color=cyan,thick] plot [mark=otimes*, mark color=cyan,samples at = {0,2}]
			({\x}, {g(\x,0.2,0.5)});
		\end{tikzpicture}%
	}
\end{lrbox}
\begin{lrbox}{\partitionA}{%
		\begin{tikzpicture}[region/.style={
				draw=black!50
			},
			Node/.style={
				midway,
			},
			declare function={
				xmin=-2;
				xmax=2;
				ymin=-2;
				ymax=2;
			},
			]
			\begin{axis}[
				xlabel={$x$},
				ylabel={$w$},
				xmin=xmin,
				xmax=xmax,
				ymin=ymin,
				ymax=ymax,
				axis background/.style={},
				extra x ticks={},
				extra y ticks={},
				xtick={},
				xticklabels={},
				yticklabels={},
				yticklabel style={color=purple},
				xtick style={draw=none},
				ytick style={draw=none}
				]
				\draw [region] (-2,-2) rectangle (-1,0)  node [Node] {$c$};
				\draw [region] (-1,-2)  rectangle (1,0)  node [Node] {$d$};
				\draw [region] (1,-2)   rectangle (2,0)  node [Node] {$e$};
				\draw [region] (-2,0)   rectangle (0.7,2) node [Node] {$a$};
				\draw [region] (0.7,0)   rectangle (2,2) node [Node] {$b$};
				%
			\end{axis}
		\end{tikzpicture}%
	}
\end{lrbox}
\begin{lrbox}{\firsttreeb}{%
		\begin{tikzpicture}
			\draw[-,scale=0.5, domain=-1.3:0, variable=\x,color=purple,thick] plot [mark=otimes*,samples at = {-1.3,0}]
			({\x}, {f(\x,0.2,0.28)});
		\end{tikzpicture}%
	}
\end{lrbox}
\begin{lrbox}{\secondtreeb}{%
		\begin{tikzpicture}
			\draw[-,scale=0.5, domain=-1:0, variable=\x,color=purple,thick] plot [mark=otimes*, samples at = {-1,0}]
			({\x}, {f(\x,0.5,0.47)});
			\draw[-,scale=0.5, domain=0:1, variable=\x,color=purple,thick] plot [mark=otimes*, samples at = {0,1}]
			({\x}, {g(\x,0.5,0.52)});
		\end{tikzpicture}%
	}
\end{lrbox}
\begin{lrbox}{\thirdtreeb}{%
		\begin{tikzpicture}
			\draw[-,scale=0.5, domain=0:2, variable=\x,color=purple,thick] plot [mark=otimes*, samples at = {0,2}]
			({\x}, {g(\x,0.2,0.67)});
		\end{tikzpicture}%
	}
\end{lrbox}
\begin{lrbox}{\fourthtreeb}{%
		\begin{tikzpicture}
			\draw[-,scale=0.5, domain=-2:0, variable=\x,color=purple,thick] plot [mark=otimes*,samples at = {-2,0}]
			({\x}, {f(\x,0.1,0.5)});
			\draw[-,scale=0.5, domain=0:2, variable=\x,color=purple,thick] plot [mark=otimes*,samples at = {0,2}]
			({\x}, {g(\x,0.1,1)});
		\end{tikzpicture}%
	}
\end{lrbox}
\begin{lrbox}{\partitionB}{%
		\begin{tikzpicture}[region/.style={
				draw=black!50
			},
			Node/.style={
				midway,
			},
			declare function={
				xmin=-2;
				xmax=2;
				ymin=-2;
				ymax=2;
			},
			]
			\begin{axis}[
				xlabel={$x$},
				ylabel={$w$},
				xmin=xmin,
				xmax=xmax,
				ymin=ymin,
				ymax=ymax,
				axis background/.style={},
				extra x ticks={},
				extra y ticks={},
				xtick={},
				xticklabels={},
				yticklabels={},
				yticklabel style={color=purple},
				xtick style={draw=none},
				ytick style={draw=none}
				]
				\draw [region] (-2,-0.6) rectangle (-0.4,2)  node [Node] {$f$};
				\draw [region] (-0.4,-0.6)  rectangle (0.4,2)  node [Node] {$g$};
				\draw [region] (0.4,-0.6)   rectangle (2,2)  node [Node] {$h$};
				\draw [region] (-2,-2)   rectangle (2,-0.6) node [Node] {$i$};
				%
			\end{axis}
		\end{tikzpicture}%
	}
\end{lrbox}
\begin{lrbox}{\partitionC}{%
		\begin{tikzpicture}[region/.style={
				draw=black!50
			},
			Node/.style={
				midway,
			},
			declare function={
				xmin=-2;
				xmax=2;
				ymin=-2;
				ymax=2;
			},
			]
			\begin{axis}[
				xlabel={$x$},
				ylabel={$w$},
				xmin=xmin,
				xmax=xmax,
				ymin=ymin,
				ymax=ymax,
				axis background/.style={},
				extra x ticks={},
				extra y ticks={},
				xtick={},
				xticklabels={},
				ytick={-0.47},
				yticklabels={$w^*$},
				yticklabel style={color=red!70!black,dashed},
				xtick style={draw=none},
				ytick style={draw=none}
				]
				\draw [region] (-2,-2) rectangle (-1,-0.6)  node [Node] {$c+i$};
				\draw [region] (-1,-2) rectangle (1,-0.6)  node [Node] {$d+i$};
				\draw [region] (1,-2) rectangle (2,-0.6)  node [Node] {$e+i$};
				\draw [region] (-2,-0.6) rectangle (-1,0)  node [Node] {$c+f$};
				\draw [region] (-1,-0.6) rectangle (-0.4,0)  node [Node] {$d+f$};
				\draw [region] (-0.4,-0.6) rectangle (0.4,0)  node [Node] {$d+g$};
				\draw [region] (0.4,-0.6) rectangle (1,0)  node [Node] {$d+h$};
				\draw [region] (1,-0.6) rectangle (2,0)  node [Node] {$e+h$};
				\draw [region] (-2,0) rectangle (-0.4,2)  node [Node] {$a+f$};
				\draw [region] (-0.4,0) rectangle (0.4,2)  node [Node] {$a+g$};
				\draw [region] (0.4,0) rectangle (0.7,2)  node [Node,rotate=90] {$a+h$};
				\draw [region] (0.7,0) rectangle (2,2)  node [Node] {$b+h$};
				\draw[color=red!70!black,dashed] (-2,-0.47) -- (2,-0.47);
				%
			\end{axis}
		\end{tikzpicture}%
	}
\end{lrbox}
\begin{lrbox}{\compositeA}{%
		\begin{tikzpicture}[region/.style={
				draw=black!50
			},
			Node/.style={
				midway,
			},
			declare function={
				xmin=-2;
				xmax=2;
				ymin=-0.4;
				ymax=2.8;
			},
			]
			\begin{axis}[
				xlabel={$x$},
				ylabel={$y$},
				axis x line=bottom,
				axis y line=left,
				axis line style={-},
				xmin=xmin,
				xmax=xmax,
				ymin=ymin,
				ymax=ymax,
				axis background/.style={},
				extra x ticks={},
				extra y ticks={},
				xtick={-1,1,-0.4,0.4},
				xticklabels={$x_1$,$x_2$,$x_3$,$x_4$},
				xtick style={draw=none}
				]
				\draw (-1,-2) to (-1,2.8);
				\draw (-0.4,-2) to (-0.4,2.8);
				\draw (0.4,-2) to (0.4,2.8);
				\draw (1,-2) to (1,2.8);
				\draw[color=gray,dashed,opacity=0.5] (0,-2) to (0,2.6);
				\node[scale=0.8] at (-1.5,2.7) {$c+f$};
				\draw[variable=\x,color=cyan] plot [mark=otimes*, mark color=black,samples at = {-1.9,-1.1}]
				({\x}, {f(\x,-0.1,-0.15)});
				\draw[variable=\x,color=purple] plot [mark=otimes*, mark color=black,samples at = {-1.9,-1.1}]
				({\x}, {f(\x,0.2,0.28)});
				\node[scale=0.8] at (-0.7,2.7) {$d+f$};
				\draw[variable=\x,color=cyan] plot [mark=otimes*, mark color=black,samples at = {-0.9,-0.5}]
				({\x}, {f(\x,0.2,0.23)});
				\draw[variable=\x,color=purple] plot [mark=otimes*, mark color=black,samples at = {-0.9,-0.5}]
				({\x}, {f(\x,0.2,0.28)});
				\node[scale=0.8] at (0,2.7) {$d+g$};
				\draw[variable=\x,color=cyan] plot [mark=otimes*, mark color=black,samples at = {-0.3,-0.1}]
				({\x}, {f(\x,0.2,0.23)});
				\draw[variable=\x,color=purple] plot [mark=otimes*, mark color=black,samples at = {-0.3,-0.1}]
				({\x}, {f(\x,0.5,0.47)});
				\draw[variable=\x,color=cyan] plot [mark=otimes*, mark color=black,samples at = {0.1,0.3}]
				({\x}, {g(\x,0.3,0.48)});
				\draw[variable=\x,color=purple] plot [mark=otimes*, mark color=black,samples at = {0.1,0.3}]
				({\x}, {g(\x,0.5,0.52)});
				\node[scale=0.8] at (0.7,2.7) {$d+h$};
				\draw[variable=\x,color=cyan] plot [mark=otimes*, mark color=black,samples at = {0.5,0.9}]
				({\x}, {g(\x,0.3,0.48)});
				\draw[variable=\x,color=purple] plot [mark=otimes*, mark color=black,samples at = {0.5,0.9}]
				({\x}, {g(\x,0.2,0.67)});
				\node[scale=0.8] at (1.5,2.7) {$e+h$};
				\draw[variable=\x,color=cyan] plot [mark=otimes*, mark color=black,samples at = {1.1,1.9}]
				({\x}, {g(\x,0.5,0.37)});
				\draw[variable=\x,color=purple] plot [mark=otimes*, mark color=black,samples at = {1.1,1.9}]
				({\x}, {g(\x,0.2,0.67)});
			\end{axis}
		\end{tikzpicture}%
	}
\end{lrbox}
\begin{lrbox}{\compositeB}{%
		\begin{tikzpicture}[region/.style={
				draw=black!50
			},
			Node/.style={
				midway,
			},
			declare function={
				xmin=-2;
				xmax=2;
				ymin=-0.4;
				ymax=2.8;
			},
			]
			\begin{axis}[
				xlabel={$x$},
				ylabel={$y$},
				axis x line=bottom,
				axis y line=left,
				axis line style={-},
				xmin=xmin,
				xmax=xmax,
				ymin=ymin,
				ymax=ymax,
				axis background/.style={},
				extra x ticks={},
				extra y ticks={},
				xtick={-1,1,-0.4,0.4},
				xticklabels={$x_1$,$x_2$,$x_3$,$x_4$},
				xtick style={draw=none}
				]
				\draw (-1,-2) to (-1,2.8);
				\draw (-0.4,-2) to (-0.4,2.8);
				\draw (0.4,-2) to (0.4,2.8);
				\draw (1,-2) to (1,2.8);
				\draw[color=gray,dashed,opacity=0.5] (0,-2) to (0,2.6);
				\node[scale=0.8] at (-1.5,2.7) {$c+f$};
				\draw[variable=\x] plot [mark=otimes*, mark color=black,samples at = {-1.9,-1}]
				({\x}, {f(\x,0.1,0.13)});
				\node[scale=0.8] at (-0.7,2.7) {$d+f$};
				\draw[variable=\x] plot [mark=otimes*, mark color=black,samples at = {-1,-0.4}]
				({\x}, {f(\x,0.4,0.51)});
				\node[scale=0.8] at (0,2.7) {$d+g$};
				\draw[variable=\x] plot [mark=otimes*, mark color=black,samples at = {-0.4,0}]
				({\x}, {f(\x,0.7,0.7)});
				\draw[variable=\x] plot [mark=otimes*, mark color=black,samples at = {0,0.4}]
				({\x}, {g(\x,0.8,1)});
				\node[scale=0.8] at (0.7,2.7) {$d+h$};
				\draw[variable=\x] plot [mark=otimes*, mark color=black,samples at = {0.4,1}]
				({\x}, {g(\x,0.5,1.15)});
				\node[scale=0.8] at (1.5,2.7) {$e+h$};
				\draw[variable=\x] plot [mark=otimes*, mark color=black,samples at = {1,1.9}]
				({\x}, {g(\x,0.7,1)});
			\end{axis}
		\end{tikzpicture}%
	}
\end{lrbox}
\begin{lrbox}{\treeA}{%
		\begin{forest}
			for tree={
				draw,
				l=15mm,
				calign=fixed edge angles,
				calign angle=40,
				align=center,
				EL/.style = {edge label={node[midway,anchor=center,fill=white,font=\tiny]{#1}},},
				LL/.style = {label={[label distance=-1mm]below:{\tiny \boxed{#1}}}},
			}
			[$w < 0$,
			[$x < 0.7$, EL=no, l sep=0pt
			[\usebox{\fifthtree},circle,LL=b,scale=0.9]
			[\usebox{\fourthtree},circle,LL=a,scale=0.7]
			]
			[$x < -1$, EL=yes
			[$x < 1$
			[\usebox{\thirdtree},circle,LL=e,scale=0.9]
			[\usebox{\secondtree},circle,LL=d,scale=0.65]
			]
			[\usebox{\firsttree}, circle,LL=c,scale=0.9]
			]
			]
		\end{forest}
	}
\end{lrbox}
\begin{lrbox}{\treeB}{%
		\begin{forest}
			for tree={
				draw,
				l=15mm,
				calign=fixed edge angles,
				calign angle=40,
				align=center,
				EL/.style = {edge label={node[midway,anchor=center,fill=white,font=\tiny]{#1}},},
				LL/.style = {label={[label distance=-1mm]below:{\tiny \boxed{#1}}}},
			}
			[$w < -0.6$,
			[$x < -0.4$, EL=no, l sep=0pt
			[$x < 0.4$,
			[\usebox{\thirdtreeb},circle,LL=h,scale=0.9]
			[\usebox{\secondtreeb},circle,LL=g,scale=0.7]
			]
			[\usebox{\firsttreeb},circle,LL=f,scale=0.9]
			]
			[\usebox{\fourthtreeb},circle,LL=i,scale=0.5,EL=yes]
			]
		\end{forest}
	}
\end{lrbox}

	\section{Simulation studies for CATE estimation in RDDs}\label{simulations}
	This section describes a parametrized protocol for simulating data for evaluating CATE estimation methods in RDDs. Modifiable code implementing this approach is available at the Github repository associated with the paper\footnote{\url{https://github.com/rafaelcalcantara/BART-RDD}}. 
	
	There are three ingredients to any simulation-based statistical method evaluation procedure: the estimand, the evaluation criteria, and the data generating process. 
	
	\subsection{The estimand}
	Generically, our estimand is the CATE function at $x = 0$; i.e. $\tau(0, \w)$. The key practical question is which values of $\w$ to consider. Some values of $\w$ will not be well-represented near $x=0$ and so no estimation technique will be able to estimate those points effectively. As such, to focus on feasible points --- which will lead to interesting comparisons between methods --- we recommend restricting the evaluation points to the observed $\w_i$ such that $|x_i| \leq \delta$, for some $\delta > 0$.  In our example, we use $\delta = 0.1$ for a standardized $x$ variable. Therefore, our estimand of interest is a vector of treatment effects:
	\begin{equation}
		\tau(0, \w_i) \;\;\; \forall i \;\mbox{ such that }\; |x_i| \leq \delta.
	\end{equation} 
	
	\subsection{Estimation loss function}	
	For our evaluation criteria we will consider average root-mean-squared estimation error, expressed as a fraction of a default ATE estimator:
	\begin{equation}
		\mbox{CATE RMSE} = \dfrac{\sqrt{\sum_{i : |x_i| \leq \delta} \left (\hat{\tau}(0, \w_i) - \tau(0, \w_i)\right )^2 }}{ \sqrt{\sum_{i : |x_i| \leq \delta} \left (\hat{\tau}(0) - \tau(0, \w_i)\right )^2 }}.
	\end{equation}
	This performance metric judges the ability of $\hat{\tau}(0, \w)$ to estimate CATEs relative to a baseline ATE estimator (at $x = 0$), thereby allowing us to tell if methods are doing better than would be possible just by assuming homogeneous effects. It also permits a unitless performance measure, so that relative accuracy across methods can be compared in a standardized way across data generating processes of varying outcome scales, which can affect the implicit difficulty of the estimation problem.

	\subsection{Data generating process}		
	
	The goal of our simulation study is to understand how various methods perform at estimating CATEs across a variety of DGPs. More particularly, we would like to be able to characterize what aspects of a DGP make a causal inference problem hard or easy so that we may identify methods which adapt to variation in the ``intrinsic'' problem difficulty.  To approach this problem we will take an analysis of variance (ANOVA) perspective \citep{hahn2018regularization, hahn2019atlantic}, tailored to the RDD context. 
	
	Because an RDD only identifies the treatment effect at $x = 0$, the relevant signal to noise ratios vis-a-vis treatment effect estimation are conditional on $x = 0$; accordingly, we will design our DGP so that it is explicitly parametrized in terms of conditional variances at $x = 0$. Data will be simulated by first generating $W$ and $X$, followed by $Y$ given $W$ and $X$.
	
	\subsubsection{Generating $(W, X)$}
	
	Our simulation studies will consider $W$ to be fixed in advance and we will consider replications over $(X, Y)$. The covariates $W$ can be empirical data from a real-world application or can be simulated. Here, for illustration purposes, we generate $W$ according to a mean-zero multivariate Gaussian distribution with a Toeplitz covariance matrix, with entries ranging from 0 to 2. For example, for $p = 5$ the covariance would be:
	$$\mathbb{C}\mbox{ov}(W) = \begin{pmatrix}
		2 &   \tfrac{3}{2}& 1 & \tfrac{1}{2} & 0 \\
		\tfrac{3}{2} & 2 & \tfrac{3}{2}& 1 & \tfrac{1}{2}\\
		1 & \tfrac{3}{2} & 2 & \tfrac{3}{2}& 1\\
		\tfrac{1}{2} &1 & \tfrac{3}{2} & 2 & \tfrac{3}{2}\\
		0 &\tfrac{1}{2} &1 & \tfrac{3}{2} & 2 
	\end{pmatrix}.$$
	We then draw $X$ according to a Gaussian distribution centered at a linear combination of the $W = \w$ values:
	$$X \mid W = \w \sim \N(\gamma_0 + \w^t \gamma, \nu)$$ where $\gamma_0$ is the marginal mean and $\gamma$ is a $p$-dimensional vector of regression coefficients. For our demonstration here we use $\gamma_0 = 1$; this choice was made so that $X$ is not centered at the cutoff, which we thought would be unrealistic. For $\gamma$ we use an evenly weighted coefficient vector such that $\mathbb{V}(X) = 1$ and $\mathbb{C}\mbox{or}(X, W^t\gamma) = \rho$, for some pre-specified value of $\rho$; these constraints also determine the value of $\nu$. Full details can be found in the Github repository associated with this paper.
	
	Setting $\gamma$ to the zero vector implies that $X$ and $W$ are independent, which is an interesting special case. But being able to test the performance of CATE estimators under varying degrees of association between the running variable $X$ and moderators $W$ is important and this linear model is a simple test case for that. To summarize, $\gamma_0$ and $\rho$ are important parameters in our DGP, governing how concentrated around the cutoff the data are and the strength of the  association between the running variable and the moderators. 		
	
	\subsubsection{Generating $Y$, given $W$ and $X$}
	
	To begin, we use the following ``treatment effect parametrization'' when specifying our DGP.
	
	\begin{equation}
		\E(Y \mid X = x, W = \w, Z = z) = \mu(x,\w) + \tau(x, \w) z,
	\end{equation}
	which relates to the notation in Section \ref{sec:rdd} by taking $\mu(x, \w) \equiv \mu_0(x, \w)$ and $\tau(x, \w) \equiv \mu_1(x, \w)-\mu_0(x, \w)$. This parametrization allows us to generate our data directly in terms of the treatment effect function; we may specify the average magnitude and complexity of $\tau(x, \w)$ explicitly. In this paper we will consider only homoskedastic errors in our DGP:
	\begin{equation}
		Y_i = \mu(x_i, \w_i) + \tau(x_i, \w_i) z_i + \sigma \epsilon_i,
	\end{equation}
	for a mean-zero Gaussian error term; extensions to heteroskedastic and/or non-normal errors are straightforward. 
	
	Before describing our specific choices for $\mu(x, \w)$ and $\tau(x, \w)$, we will discuss a strategy for fixing some properties of these functions averaged over $W$, given $X = 0$. Specifically, we consider the following properties/quantities: 
	$\min_{\w} \tau(0, \w)$, $\V(\tau(0, W) \mid X = 0)$, and $\V(\mu(0, W) \mid X = 0)$. We use Monte Carlo simulation to compute these quantities for ``template'' functions $\mu^{\star}$ and $\tau^{\star}$ and then devise linear transformations of those template functions to achieve desired relationships between them. That is, we take a large sample from $W \mid X = 0$ and compute the above quantities based on that simulated data. For $(W, X)$ draw as described above, $W \mid X = 0$ is a multivariate Gaussian with 
	\begin{equation}
		\begin{split}
			\E(W \mid X = 0) &= - \Sigma_W \gamma \gamma_0,\\
			\V(W \mid X = 0) &= \Sigma_W - \Sigma_W \gamma \gamma^t \Sigma_W^t.
		\end{split}
	\end{equation}
	Using this strategy, we fix $\V(\mu(0, W) \mid X = 0) = 1$ and specify our DGP in terms of the following parameters
	\begin{equation}
		\begin{split}
			\sqrt{\V(\tau(0, W) \mid X = 0)} & = k_2,\\
			\sqrt{\V(Y \mid X = 0, W)} = \sigma &= k_4\\
			\min_{\w} \tau(0, \w) = \ubar{\tau}_0 &= k_5.
		\end{split}
	\end{equation}
	These values, along with $\gamma_0$ and $\rho$ mentioned in the previous section, are the key parameters in our DGP.
	
	Last, but not least, we must specify $\mu^{\star}$ and $\tau^{\star}$.	Letting $\w = (w_1 \dots w_p)$ be realizations of a length $p$ random vector $W$, define $w^{\star} = \frac{\sum_{j=1}^p w_j}{\sqrt{p}}$. Our template functions are:
	\begin{equation}
		\begin{split}
			\mu^{\star}(x,\w) &= k_1 (x+1)^3 + (w^{\star}+2)^2 \left ( \text{sign}(x+1) \sqrt{|(x+1)|} \right )^{k_3},\\
			\tau^{\star}(\w) &= \Phi(2 w_1 + 3)/2 + \phi(w_1),
		\end{split}
	\end{equation}
	where $\Phi(\cdot)$ and $\phi(\cdot)$ are the cumulative distribution and probability density functions, respectively, of a standard normal random variable. The variables $k_1$ and $k_3$ are further parameters for variation; the parameter $k_1$ controls how much variability of $\mu$ will be due to $X$ versus to $W$ and $k_3$ determines whether or not $\mu$ is additive in $X$ and $W$ or if there is an interaction ($k_3 = 1$) or not ($k_3 = 0$). 
	
	These choices of $\mu^{\star}$ and $\tau^{\star}$ provide nontrivial nonlinearities while being relatively easy to understand. Furthermore, they are designed such that the dimension of $W$ can be modified without changing the function definition, but still using all available dimensions in the definition of $\mu$. The treatment effect function is restricted to depend on a single element of $W$ to facilitate plotting.
	
	Many variations of these functions were explored in the preparation of this paper; these specific choices nicely illustrate the simulation procedure.
	Considering a variety of template functions is of course recommended and should be chosen depending on domain knowledge to investigate their impact on the performance of candidate CATE estimation procedures in a use-case-relevant way.
	
	\subsection{Estimation methods}
	To demonstrate our simulation protocol we will compare the following methods:
	
	\begin{itemize}
		\setlength{\itemsep}{-1.5em}
		\setlength{\parskip}{0pt}
		\setlength{\parsep}{0pt}
		\item BARDDT\\
		\item S-BART\\
		\item T-BART\\
		\item  local polynomial estimators (with and without regularization)\\
		\item RD-Tree \citep{reguly2021heterogeneous}.
	\end{itemize}
	All three BART variants were fit with 50 trees each (two forests of 50 trees for T-BART), with tree depth parameters set as in \cite{chipman2010bart}: $\alpha=0.95,\beta=2$ and fit using the {\tt stochtree} package. Further, the CATE estimator in all cases was the vector of posterior means of $\tau(0, \w_i)$ for $i$ such that $|x_i| \leq 0.1$.
	
	The local polynomial estimator is trained on data points within the bandwidth obtained with the {\tt rdrobust} package \citep{calonico2015rdrobust}. Ordinary-least-squares is used to fit a fourth degree polynomial in each feature of $W$, interacted with $X$ and $Z$, or, in Wilkinson notation \citep{wilkinson1973symbolic}:
	\begin{equation}
		Y \sim \left( \sum_{j=1}^p \text{poly}(W_j,4) \right) \cdot X \cdot Z ,
	\end{equation}
	where $\text{poly}(A,p) = \sum_{j=1}^p A^j$. The merely-linear specification in the running variable is consistent with the approach of \cite{calonico2025treatment}; as the bandwidth parameter approaches zero, only a linear term is required to estimate the treatment effects at the cutoff.  The choice of this particular polynomial (in $W$) is of course open to discussion, but any specific choice will likewise be subject to critique; a main benefit of tree-based regressions is that they side-step this decision. To fit this model we use both ordinary least squares (OLS), as well as two regularized Bayesian specifications, one using a conjugate normal prior (i.e., ridge regression) and one using the horseshoe prior \citep{carvalho2009handling}. We use the {\tt bayeslm} package to fit these Bayesian linear models \citep{hahn2019efficient}.
	
	Model parameters for RD-Tree were set as suggested in \citet{reguly2021heterogeneous}. For further details on the method and its parameters, please refer to the original text and the example script at \url{https://github.com/regulyagoston/RD_tree/}.

	\subsection{Comparisons}	
	
	The results in this section are based on configurations of the DGP described in the previous section which can be roughly separated into two groups: ``easy'' and ``hard''. For the easy setting, $(k_1 = 1,k_2 = 1,k_3 = 0,k_4 = 0.1)$: prognostic and treatment variation are comparable magnitudes, $\mu$ is separable in $x$ and $\w$, and low noise. For the ``hard'' setting $(k_1 = 5 ,k_2 = 0.25 ,k_3 = 1, k_4 = 0.5)$: prognostic variation is twenty times larger than treatment variation, $\mu$ is non-separable in $x$ and $\w$, and noise is high. Table \ref{tab:dgps} presents the parameters for each DGP configuration. The `easy' setting corresponds to the first 3 rows of table 1; the `hard' settings correspond to the last 3 rows. Results are based on 100 replications of size $n=4000$ for each DGP configuration. 
	
	\begin{table}[ht]
	\centering
	\small
	\begin{tabular}{cccccccc}
		\hline
		DGP & $k_1$ & $k_2$ & $k_3$ & $k_4$ & $k_5$ & $p$ & $\rho$\\
		\hline
		1 & 1.00 & 1.00 & 0.00 & 0.10 & 0.00 & 2.00 & 0.50\\ 
		2 & 1.00 & 1.00 & 0.00 & 0.10 & 0.00 & 4.00 & 0.00 \\ 
		3 & 1.00 & 1.00 & 0.00 & 0.10 & 1.00 & 2.00 & 0.00 \\ 
		4 & 5.00 & 0.25 & 1.00 & 0.50 & 0.00 & 4.00 & 0.50\\ 
		5 & 5.00 & 0.25 & 1.00 & 0.50 & 1.00 & 2.00 & 0.50\\ 
		6 & 5.00 & 0.25 & 1.00 & 0.50 & 1.00 & 4.00 & 0.00\\ 
		\hline
	\end{tabular}
	\caption{\small DGP configurations investigated in the simulations. The settings can be roughly separated into two groups: ``easy'' and ``hard''. For the easy setting, $(k_1 = 1,k_2 = 1,k_3 = 0,k_4 = 0.1)$: prognostic and treatment variation are comparable magnitudes, $\mu$ is separable in $x$ and $\w$, and low noise. For the ``hard'' setting $(k_1 = 5 ,k_2 = 0.25 ,k_3 = 1, k_4 = 0.5)$: prognostic variation is twenty times larger than treatment variation, $\mu$ is non-separable in $x$ and $\w$, and noise is high. The first 3 rows correspond to the `easy' setting; the last 3 rows correspond to the `hard' settings.} 
	\label{tab:dgps}
\end{table}
	

	\subsubsection{Overall results}	
\begin{figure}[!htpb]
	\centering
	\includegraphics[width=0.95\linewidth]{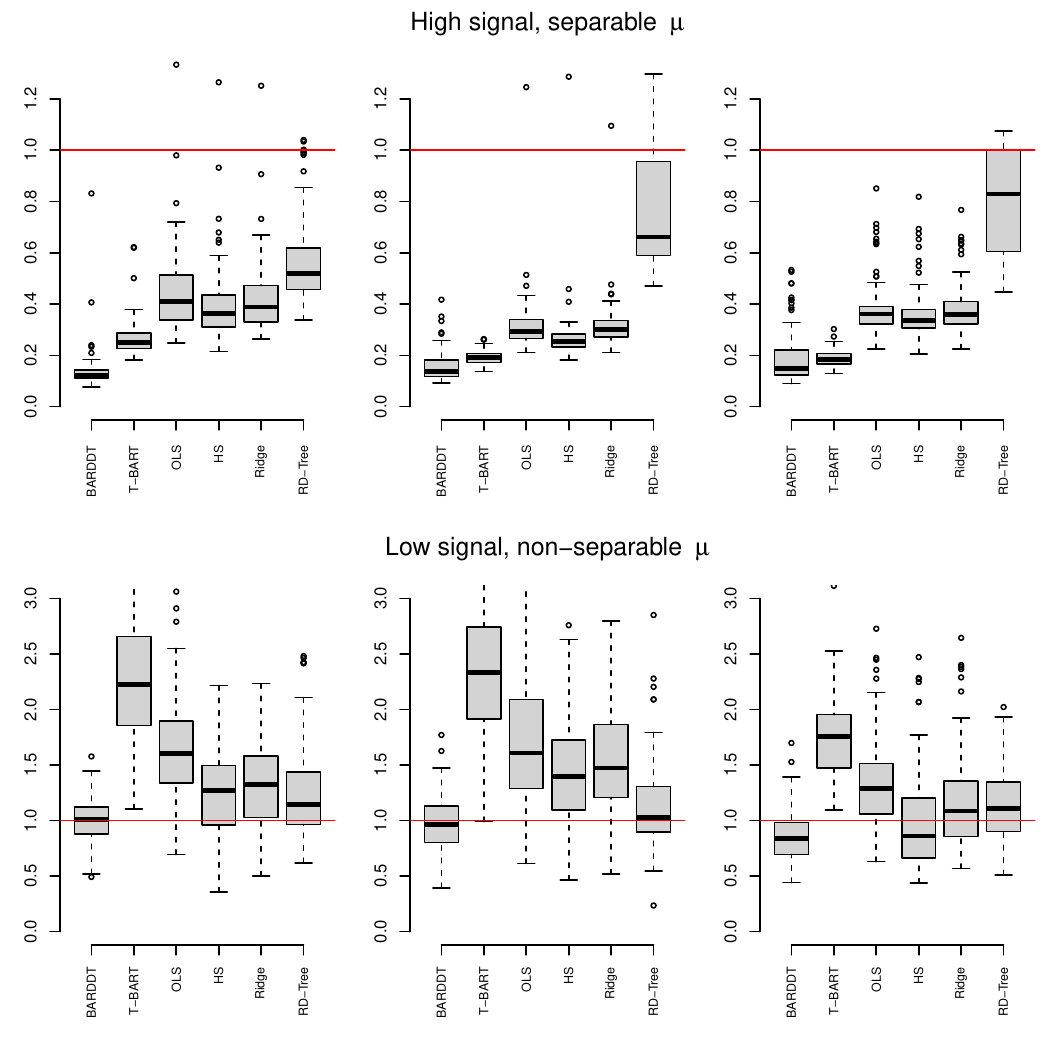}
	\caption{\small RMSE results per estimator. The upper row corresponds to the `easy' setup described in the text; the lower row corresponds to the `hard' setup. For the easy setting, the values for the additional parameters are $(k_5,p,\rho)=\{(0,2,0.5),(0,4,0),(1,2,0)\}$; for the hard setting, the values are $(k_5,p,\rho)=\{(0,4,0.5),(1,2,0.5),(1,4,0)\}$. The CATE RMSE values are divided by the RMSE obtained by predicting the CATE with an estimated ATE. Thus, the red line at 1 indicates the point above which the methods are worse than this naive estimator. BARDDT produces lower RMSE than the other estimators on average and is most often better than the homogeneous estimator.}
	\label{fig:boxplots}
\end{figure}

	Figure \ref{fig:boxplots} shows boxplots of the CATE RMSE per replication for each one of the ``easy'' and ``hard'' setups. S-BART results are excluded from these plots because it had significantly worse performance in all scenarios, distorting the plotting scale and making comparing the other methods more difficult visually. The RMSE values are normalized by dividing through by the RMSE that would be obtained by predicting the CATE with an estimated ATE, which we obtained using the {\tt rdrobust} package. This means that a value greater than or equal to 1 means the estimator is doing no better at CATE prediction than estimating the ATE and assuming treatment effect homogeneity. BARDDT produces lower RMSE than the other estimators on average and is most often better than the homogeneous estimator.
	
	Table \ref{tab:results} presents the average RMSE obtained by all estimators. BARDDT is always better than the other methods. T-BART performs well in the easier settings, but suffers a near tenfold increase in its average RMSE in the settings with high noise and non-separable $\mu$. The local polynomial estimator usually performs worse than BARDDT, even in the settings in which it is competitive on average, and it frequently obtains much larger errors in individual replications. RD-Tree is competitive with the polynomial for second-best in the harder DGPs, but is the second worst performer in the easier cases.
	
\begin{table}[ht]
	\centering
	\begin{tabular}{cccccccc}
		\hline
		DGP & BARDDT & T-BART & S-BART & OLS & HS & Ridge & RD-Tree \\ 
		\hline
		1 & 0.14 & 0.27 & 1.18 & 0.44 & 0.40 & 0.42 & 0.57 \\ 
		2 & 0.16 & 0.19 & 1.50 & 0.31 & 0.27 & 0.32 & 0.74 \\ 
		3 & 0.19 & 0.19 & 1.66 & 0.38 & 0.36 & 0.38 & 0.80 \\ 
		4 & 1.01 & 2.26 & 1.60 & 1.66 & 1.23 & 1.30 & 1.30 \\ 
		5 & 0.98 & 2.34 & 3.53 & 1.72 & 1.45 & 1.56 & 1.11 \\ 
		6 & 0.85 & 1.75 & 2.55 & 1.33 & 0.99 & 1.17 & 1.14 \\ 
		\hline
	\end{tabular}
	\caption{Average RMSE per DGP, also divided here by the RMSE of the naive ATE estimator} 
	\label{tab:results}
\end{table}
	
	\subsubsection{Individual fits}	
\begin{figure}
	\centering
		\includegraphics[width=0.85\textwidth]{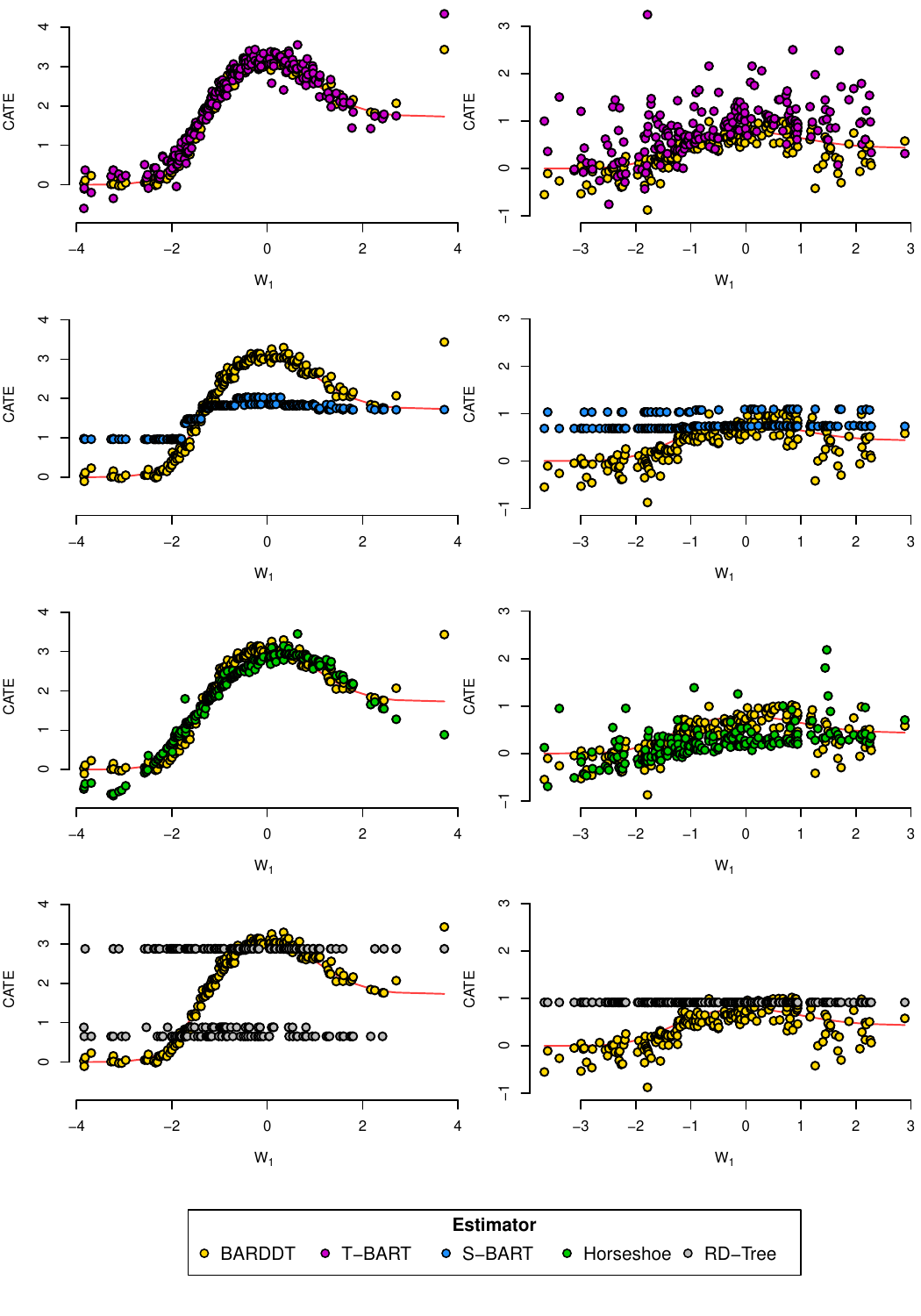}
	\caption{\small Each panel presents the CATE fits for one illustrative sample for one easy and one hard DGP setting, plotted against the true CATE, shown as a red curve. Left panel corresponds to the parameter configuration in the first row of table \ref{tab:results}, right panel corresponds to the fourth row. The BARDDT fit is shown in gold in all of the plots for ease of comparison. From top to bottom we have T-BART (purple), S-BART (blue), linear model with horseshoe prior (green), and RD-Tree (gray).}
	\label{fig:fits}
\end{figure}

	In addition to the aggregate results, such as those reported in Table \ref{tab:results} and depicted in Figure \ref{fig:boxplots}, it is often instructive to consider individual fits compared to the ground-truth, which is available to us in simulation studies. The qualitative behavior --- the specific ways that the models misfit certain types of data --- is often persistent across replications, but can be seen in individual fits. Figure \ref{fig:fits} presents the CATE fits for an individual data set, for one easy and one hard DGP. The qualitative fits of BARDDT are in line with expectations, while the other methods exhibit undesirable behavior in certain cases.
	
	\begin{itemize}
		\item Although T-BART performs well under the easier regime, even there it still exhibits high variance CATE estimates. T-BART's high variance becomes more pronounced under the harder DGP, resulting in substantially higher RMSE relative to both BARDDT and the polynomial model\\
		\item The extreme bias shift exhibited by S-BART in the low noise setting is reminiscent of the regularization-induced confounding (RIC) problem, described by \cite{hahn2020bayesian}. Broadly, the lesson here is that S-BART has unpredictable biases in causal inference problems. It does comparatively well in the high noise case, but only because it rarely splits in that case, collapsing to a homogeneous treatment model, which outperforms the overfitting T-BART and polynomial models in this regime\\
		
		\item The fits for the easier setup show that, even with high signal, the polynomial model struggles with extrapolation at the boundaries of the support of $w_1$. At the same time, the polynomial model also presents a sizable increase in variance under high noise, as seen on the fits for the harder regime\\
		
		\item RD-Tree appears to ``under-split'' on $W$, leading to a too-coarse fit of the CATE function, especially in the low-noise regime. This behavior is to be expected with a single CART fit, a problem that additive tree models, like BART, were explicitly designed to address
	\end{itemize}
	
	\subsection{Comparison to commonly studied simulation protocols}
	Before moving on, it is important to discuss how our simulation protocol differs from common practice for simulation studies in the methodological RDD literature. The most common setup comes from \cite{imbens2012optimal}, who generate data with a mean function that is a polynomial of $X$, with coefficients determined by a polynomial fit of the election data in \cite{lee2008randomized}. \cite{calonico2014robust} consider additionally a setup based on \cite{ludwig2007does}, where again the mean function is a polynomial of $X$ with coefficients determined by a polynomial fit of the original data.
	
	Simulations based on at least one of these two setups have originally been used for investigating RDD estimators for the local ATT --- or confidence intervals for such estimators --- which did not consider additional covariates (some examples besides \cite{imbens2012optimal} and \cite{calonico2014robust} include \cite{branson2019nonparametric,calonico2020optimal}). Although one might argue that, for this particular context, this is not problematic since high-degree polynomials can, in principle, be used to approximate arbitrarily complex univariate functions, the picture is much more complicated when we consider multivariate mean functions. Studies that develop estimators targeting the local ATT which use additional covariates for precision gains typically consider variations of these setups with extra covariates entering the mean function additively and with no interactions with $X$. Some examples of this approach include \cite{chib2023nonparametric,calonico2019regression,frolich2019including,kreiss2023inference}. \cite{reguly2021heterogeneous} provides the only other simulation protocol for a RDD CATE estimator that we are aware of. In that study, there is a variation of the \cite{lee2008randomized} setup with a binary $W$ determining different coefficients for the $X$ polynomial, and a variation of the \cite{ludwig2007does} setup in which a continuous $W$ is included linearly and only in the treatment effect function.
	
	Our protocol not only allows for nontrivial nonlinearities in the DGP, but also takes into account other relevant features for RDD estimation that are overlooked by the commonly used protocols. First of all, we consider different combinations of prognostic/treatment effect variation and noise levels, which means we can investigate our method under both easy and hard scenarios. In contrast, other studies typically not only ignore this kind of variation, but set an implausibly high signal. Beyond that, we consider the signal and noise levels \emph{locally at the cutoff}, rather than globally, which is what truly matters in order to determine how hard an RDD estimation problem will be. We also explore models with and without interactions between $W$ and $X$, and with more complicated interaction patterns than the few studies that do consider interactions. Finally, we also consider different levels of dependence between $W$ and $X$, an important feature that is often overlooked. In summary, our protocol not only encompasses the most commonly used protocols, but allows for a much richer exploration of estimator properties that is not tied to a particular modelling assumption and that can be easily adapted to study estimator properties in more specific use cases.
	
	\section{The Effect of Academic Probation on Educational Outcomes}
	\label{sec:application}
	We turn now to an empirical illustration based on \citet{lindo2010ability}, who analyze data on college students enrolled in
	a large Canadian university in order to evaluate the effectiveness of an academic probation policy. Students who present a grade point average (GPA) lower than a certain threshold at the end of each term are placed on
	academic probation and must improve their GPA in the subsequent
	term or else face suspension. We are interested in how being put on probation or not, $Z$, affects students' GPA, $Y$, at the end of the current term. The running variable, $X$, is the negative distance between a student's previous-term GPA and the
	probation threshold, so that students placed on probation ($Z = 1$) have a
	positive score and the cutoff is 0. Potential moderators, $W$, are:
	\begin{itemize}
		\setlength{\itemsep}{0.5em}
		\setlength{\parskip}{0pt}
		\setlength{\parsep}{0pt}
		\item  gender (`male'), 
		\item age upon entering university (`age\_at\_entry')
		\item a dummy for being born in
		North America (`bpl\_north\_america'), 
		\item the number of credits taken in the first year (`totcredits\_year1')
		\item an indicator designating each of three campuses (`loc\_campus' 1, 2 and 3), and
		\item high school GPA as a quantile w.r.t the university's incoming class (`hsgrade\_pct').
	\end{itemize}

\begin{figure}[!htpb]
	\centering
		\includegraphics[width=0.65\textwidth]{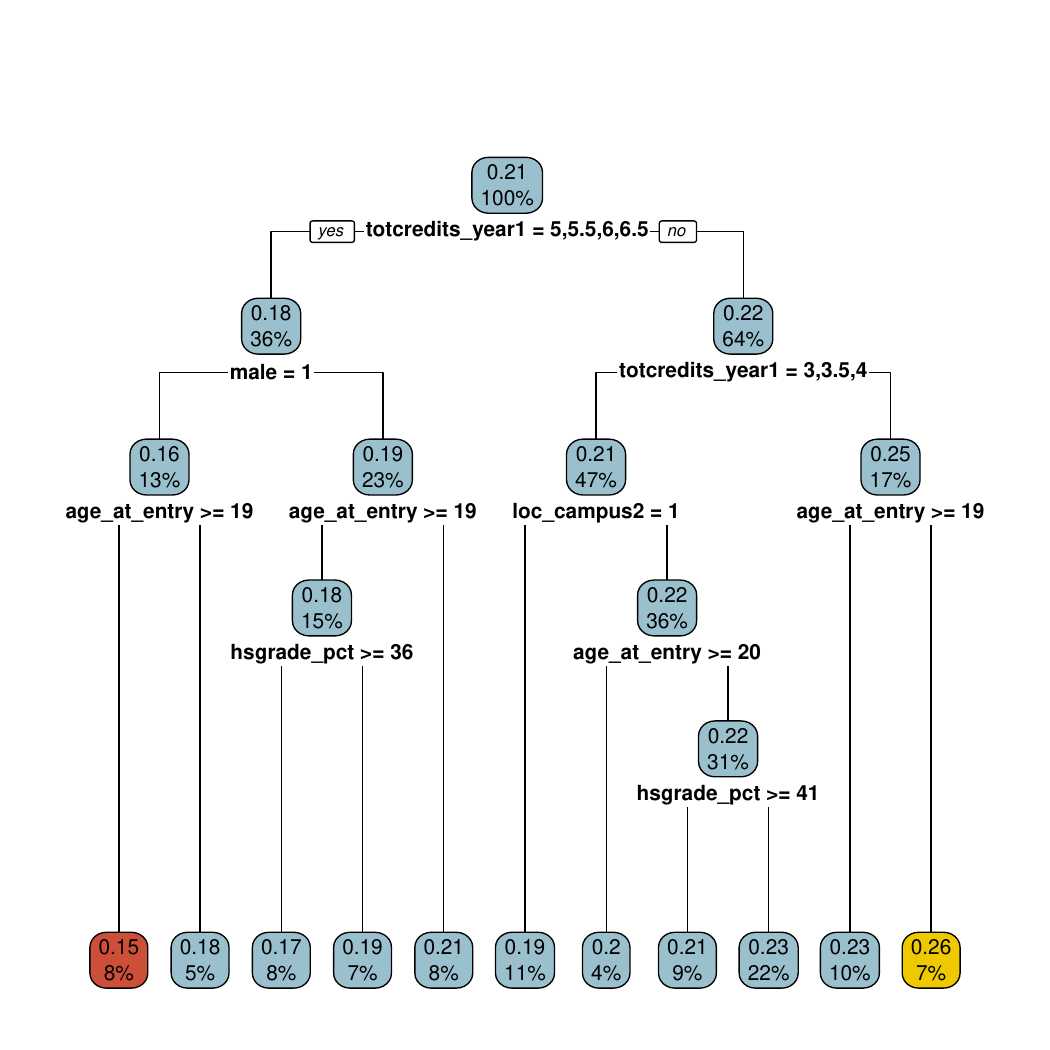}
		\caption{\small Regression tree fit to posterior point estimates of individual treatment effects: top number in each box is the average subgroup treatment effect, lower number shows the percentage of the total sample in that subgroup; the tree flags credits in first year, gender, and age at entry as important moderators}
		\label{fig:posterior.cart}
		\includegraphics[width=0.75\textwidth]{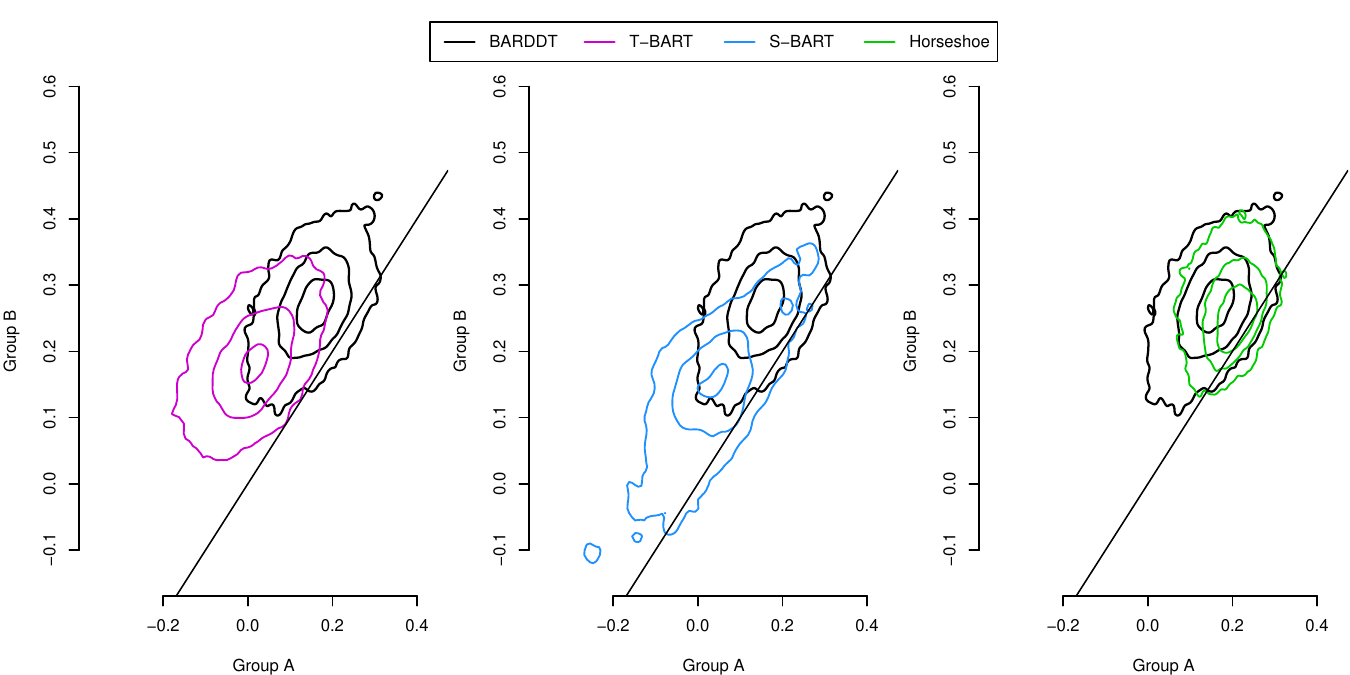}
			\caption{\small Kernel density estimates for the joint CATE posterior between male students who entered college older than 19 and attempted at least 5 credits in the first year (leftmost leaf node, red) and students who entered college younger than 19 and attempted between 4 and 5 credits in the first year (rightmost leaf node, gold). Black contour lines show the joint CATE posterior obtained by BARDDT, which is contrasted with the joint posterior obtained by T-BART, S-BART and horseshoe, respectively}
		\label{fig:posterior.difference}
\end{figure}

	Figure \ref{fig:posterior.cart} presents a summary of the CATE posterior produced by BARDDT for this application. This picture is produced fitting a regression tree, using $W$ as the predictors, to the individual posterior mean CATEs:
	\begin{equation}
		\bar{\tau}_i =  \frac{1}{M} \sum_{h = 1}^M \tau^{(h)}(0, \w_i),
	\end{equation}
	where $h$ indexes each of $M$ total posterior samples.  As in our simulation studies, we restrict our posterior analysis to use $\w_i$ values of observations with $|x_i| \leq \delta = 0.1$ (after normalizing $X$ to have standard deviation 1 in-sample). For the  \citet{lindo2010ability} data, this means that BARDDT was trained on $n = 40,582$ observations, of which 1,602 satisfy $|x_i| \leq 0.1$, which were used to generate the effect moderation tree from Figure \ref{fig:posterior.cart}.
	
	The resulting effect moderation tree indicates that course load (credits attempted) in the academic term leading to their probation is a strong moderator. Contextually, this result is plausible, both because course load could relate to latent character attributes that influence a student's responsiveness to sanctions and also because it could predict course load in the current term, which would in turn have implications for the GPA (i.e. it is harder to get a high GPA while taking more credit hours).  The tree also suggests that effects differ by campus, and age and gender of the student. These findings are all prima facie plausible as well. 
	
	To gauge how strong these findings are statistically, we can zoom in on isolated subgroups and compare the posteriors of their subgroup average treatment effects. This approach is valid because in fitting the effect moderation tree to the posterior mean CATEs we in no way altered the posterior itself; the effect moderation tree is a posterior summary tool and not any additional inferential approach; the posterior is obtained once and can be explored freely using a variety of techniques without vitiating its statistical validity. Investigating the most extreme differences is a good place to start: consider the two groups of students at opposite ends of the treatment effect range discovered by the effect moderation tree:	
	\begin{itemize}
		\setlength{\itemindent}{0.5in}
		\item[{\bf Group A}] a male student that entered college older than 19 and attempted at least 5 credits in the first year (leftmost leaf node, colored red,  comprising 128 individuals)
		\item[{\bf Group B}] a student of any gender who entered college younger than 19 and attempted more than 4, but less than 5 credits in the first year (rightmost leaf node, colored gold, comprising 108 individuals). 
	\end{itemize}
	
	Subgroup CATEs are obtained by aggregating CATEs across the observed $\w_i$ values for individuals in each group; this can be done for individual posterior samples, yielding a posterior distribution over the subgroup CATE:
	\begin{equation}
		\bar{\tau}_A^{(h)} = \frac{1}{n_A} \sum_{i : \w_i} \tau^{(h)}(0, \w_i),
	\end{equation}
	where $h$ indexes a posterior draw and $n_A$ denotes the number of individuals in group A.
	Figure \ref{fig:posterior.difference} presents a contour plot for a bivariate kernel density estimate of the joint CATE posterior distribution for subgroups A and B. The contour lines are almost all above the $45^{\circ}$ line, indicating that the preponderance of posterior probability falls in the region where the treatment effect for Group B is greater than that of Group A, meaning that the difference in the subgroup treatment effects flagged by the effect moderation tree persists even after accounting for estimation uncertainty in the underlying CATE function.
	
	Here we can compare the results for various methods. Specifically, each of the three panels shows the bivariate kernel density estimates of BARDDT compared to T-BART, S-BART, and local polynomial horseshoe, respectively. In this instance, local polynomial horseshoe is quite similar to BARDDT, although with reduced uncertainty. These comparisons constitute a sensitivity analysis, although in light of the simulation results from above, we argue that the BARDDT results should be favored. 
	
	As always, CATEs that vary with observable factors do not necessarily represent a {\em causal} moderating relationship. Here, if the treatment effect of academic probation is seen to vary with the number of credits, that does not imply that this association is causal: prescribing students to take a certain number of credits will not necessarily lead to a more effective probation policy, it may simply be that the type of student to naturally enroll for fewer credit hours is more likely to be responsive to academic probation. Similarly, even though the tree singles out students who took exactly 4.5 credits as those for whom the treatment is most effective, we have no reason \textit{a priori} to believe there is anything about that particular course load which increases effectiveness of the probation policy. An entirely distinct set of causal assumptions is required to interpret the CATE variations themselves as causal. All the same, uncovering these patterns of treatment effect variability is crucial to suggesting causal mechanisms to be investigated in future studies. For example, the results suggest it would be useful to collect more data on student and course characteristics in order to understand whether the types of students who take up to 4 credits, at least 5 credits or exactly 4.5 credits differ in ways that could affect their responsiveness to treatment.
	
	We close this section by noting the similarities and a few notable differences between our results and those of \citet{lindo2010ability}. First, we note that we find gender and high-school grades to correlate with treatment effects in the same direction as \citet{lindo2010ability} (\textit{i.e.} larger effect for women and for lower high-school grade percentiles). The important distinction is that we find that these features interact with age and course load in our moderation analysis. Additionally, the authors do not explore course load and age at all as potential moderators, which our results strongly suggest. On the other hand, the authors find evidence for variation in treatment effects by birth place, which our posterior summary tree does not indicate.
	
	\subsection{Scalability to high-dimensional settings}
\begin{figure}[!htpb]
	\centering
	\includegraphics[scale=0.7]{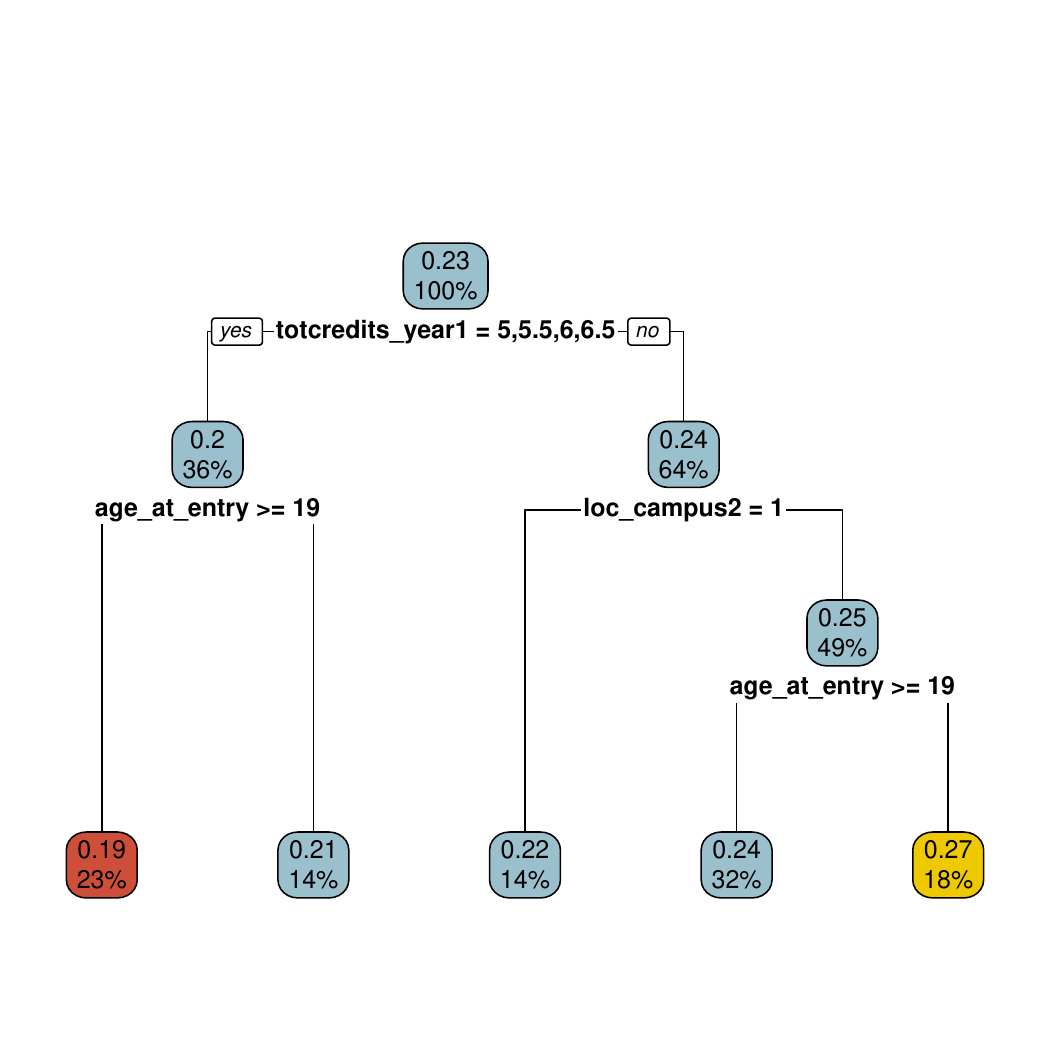}
	\caption{\small CART summary tree for BARDDT fit of the probation data with added null features. We created 20 $\mathcal{N}(0,1)$ features and added them in the estimation. The CART fit here looks very similar as the original one, picking most of the same features as important, and producing similar CATE estimates for these groups}
	\label{fig:added.features}
\end{figure}

	In order to investigate the scalability of our model in terms of number of variables, we ran the empirical analysis again with added null features. Specifically, we generated 20 Gaussian features and added them to the problem, without alerting the algorithm to the fact that they are, in fact, useless. What we observe is that the results are virtually the same as in the original analysis, demonstrating that the BARDDT method can cope with this extra covariates effectively. Figure \ref{fig:added.features} presents the CART summary of this new fit, similar to figure \ref{fig:posterior.cart}. We see that the variables picked out in the tree are essentially the same as before, with none of the null features being selected. The CATE estimates within the CART leaves are also essentially the same as before. Table \ref{tab:null.features.correlation} shows that the correlation between our CATE predictions and each of the null features is effectively zero. The results presented here speak to the ability of BARDDT to handle more features. Finally, we note that including more than a few dozen variables is feasible, but would require a very large sample size for proper estimation at the boundaries of the cutoff.
	
	\begin{table}[ht]
\centering
\begin{tabular}{lc}
& Correlation \\
\hline
$W_{1}$ & -0.014 \\
$W_{2}$ & -0.028 \\
$W_{3}$ & 0.019 \\
$W_{4}$ & 0.004 \\
$W_{5}$ & -0.012 \\
$W_{6}$ & 0.012 \\
$W_{7}$ & 0.016 \\
$W_{8}$ & -0.030 \\
$W_{9}$ & 0.009 \\
$W_{10}$ & -0.009 \\
$W_{11}$ & 0.024 \\
$W_{12}$ & 0.041 \\
$W_{13}$ & 0.012 \\
$W_{14}$ & 0.015 \\
$W_{15}$ & -0.045 \\
$W_{16}$ & 0.014 \\
$W_{17}$ & 0.015 \\
$W_{18}$ & 0.004 \\
$W_{19}$ & -0.039 \\
$W_{20}$ & 0.010 \\
\hline
\end{tabular}
\caption{\small Correlation between the treatment effect posterior mean and noise features}
\label{tab:null.features.correlation}
\end{table}

	
	
	
	\section{Summary}
	Reliable CATE estimation is important for making the most of our observational data sets. As RDD continues to gain popularity in industry --- for example, as a byproduct of business decisions being made based on an observed index --- being able to use these data to explore subgroup treatment effects is a big advantage. In this paper, we have demonstrated that a BART ensemble of treed linear regressions --- which we call BARDDT --- estimates RDD CATEs successfully and markedly better than available alternatives and have demonstrated how to interpret the resulting estimates on a reanalysis of a policy evaluation question from education \citep{lindo2010ability}. Software for fitting BARDDT is freely available in the {\tt stochtree} package, available in both {\tt R} and Python.  
	
	Recall that previous non-Bayesian CATE work for RDDs \citep{becker2013absorptive, calonico2025treatment} assumes a known basis expansion in the covariates (other than the running variable). In doing so, they achieve ``linearity in the parameters'', which in turn facilitates frequentist theoretical analysis. However, this approach avoids the fundamental challenge of nonlinear CATE estimation without really solving it -- finding a suitable basis is at least half the battle. The BART model developed in this paper does not admit the sort of frequentist theoretical analysis long-favored in mainstream econometrics, but it does provide a rigorous Bayesian treatment of the problem, bringing to bear an effective modern computational strategy (tree ensembles) for nonlinear CATE estimation, rather than merely assuming the problem away.
	
	Likewise, recall that previous Bayesian approaches to CATE estimation in RDDs either 
	\begin{itemize}
		\item	do not permit CATE estimation \citep{chib2023nonparametric}, or
		\item are computationally prohibitive in practice \citep{karabatsos2015bayesian, branson2019nonparametric}, or
		\item require prior knowledge of the heterogeneous subgroups  \citep{sugasawa2023hierarchical, tao2025bayesian}.
	\end{itemize}	
	BARDDT solves all of these problems, permitting data-driven CATE discovery on realistically large data sets. 
	
	Additionally, Section \ref{simulations} lays out desiderata for designing data generating processes that are realistic, but for which CATE estimation is feasible in principle. The proposed family of DGPs addresses a notable flaw in the existing literature, which is that CATE estimation procedures are typically demonstrated using DGPs that consist of overly-simplistic response surfaces with improbably large treatment effects and implausibly conspicuous heterogeneity. By contrast, our family of DGPs accommodates complex response surfaces while allowing direct control over the variation and magnitude of the implied CATEs. Designing test DGPs along the lines of the protocol described here will allow good-faith comparisons of future innovations in heterogeneous treatment effect estimation in RDDs.	

	\newpage
	\bibliography{alcantara_et_al}
	\newpage
	\appendix
	\section{Prior sensitivity}
\begin{figure}[!htpb]
	\centering
	\includegraphics[scale=0.7]{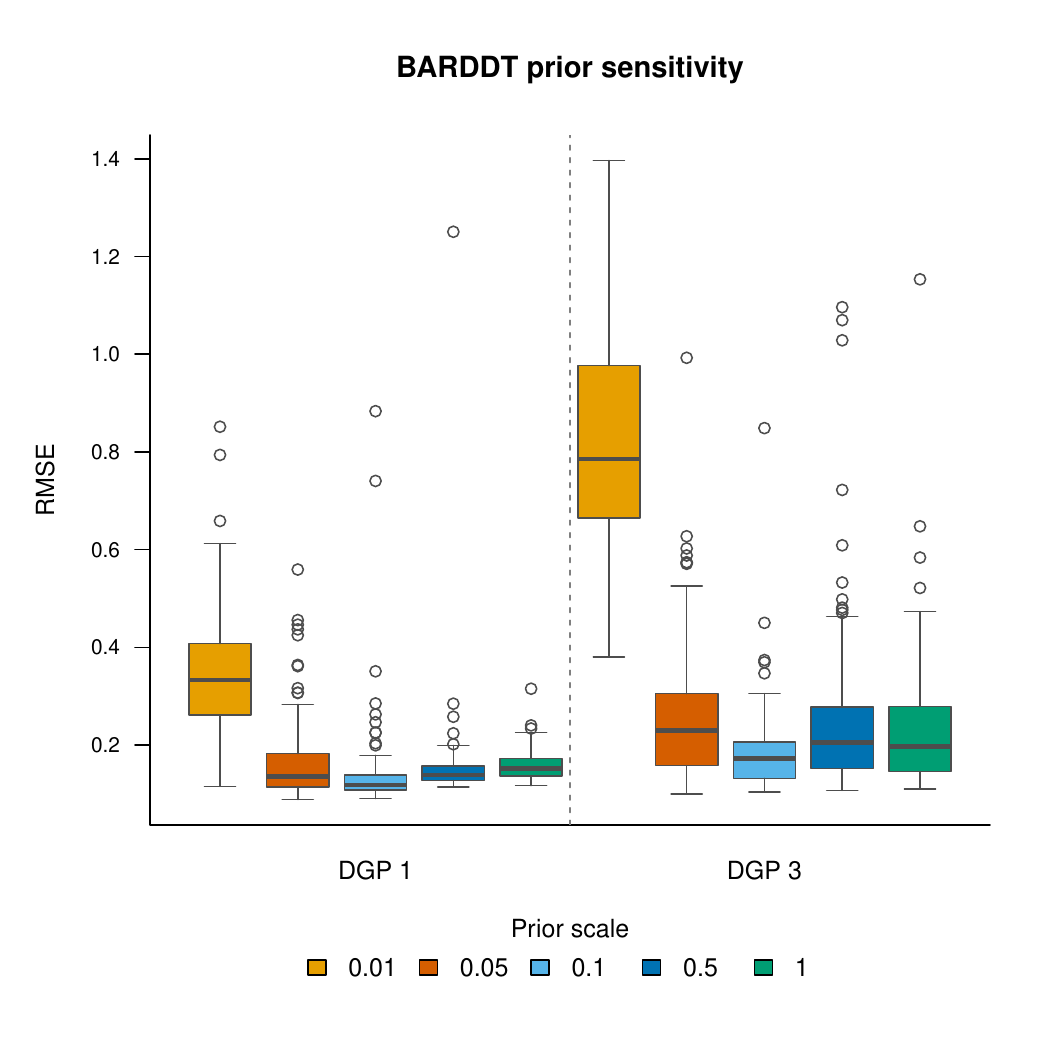}
	\caption{\small Comparison of BARDDT results for DGPs 1 and 3 in the simulations for different scales on the leaf regression coefficient priors. Results are consistent with the ones presented in the main text except for the case when the scale is very low (0.01)}
	\label{fig:prior.sensitivity}
\end{figure}

	In the process of developing our simulation study, we found that the scale of the regression coefficient priors does not affect the results dramatically unless it is set unreasonably low. We provide a quick illustration of this below; the settings considered here correspond to DGPs 1 and 3 in our simulations. We consider the following values for the prior scale: $$\begin{bmatrix}
		\frac{0.01}{J}I & \frac{0.05}{J}I & \frac{0.1}{J}I & \frac{0.5}{J}I & \frac{1}{J}I
	\end{bmatrix}.$$
	The results in figure \ref{fig:prior.sensitivity} indicates that our simulation results are not driven by particular prior choices, and that the method works well with relatively vague priors.
\end{document}